\documentclass[twocolumn]{aastex631}

\usepackage{CJKutf8} 

\usepackage{amsmath}

\newcommand\omc{$\omega$\,Cen}

\defcitealias{2023ApJ...958....8N}{Paper I}
\defcitealias{2024ApJ...970..192H}{Paper II}
\defcitealias{2015ApJ...803...29W}{Watkins15}

\usepackage{xurl}

\received{January 16, 2025}
\revised{March 4, 2025}
\accepted{March 5, 2025}

\begin{document}
\begin{CJK*}{UTF8}{gbsn}
\title{oMEGACat. VI. Analysis of the overall kinematics of Omega Centauri in 3D: velocity dispersion, kinematic distance, anisotropy, and energy equipartition}

\correspondingauthor{Maximilian Häberle}
\email{haeberle@mpia.de}
\author[0000-0002-5844-4443]{Maximilian Häberle}
\affiliation{Max Planck Institute for Astronomy, K\"onigstuhl 17, D-69117 Heidelberg, Germany}
\author[0000-0002-6922-2598]{N. Neumayer}
\affiliation{Max Planck Institute for Astronomy, K\"onigstuhl 17, D-69117 Heidelberg, Germany}
\author[0009-0005-8057-0031]{C. Clontz}
\affiliation{Max Planck Institute for Astronomy, K\"onigstuhl 17, D-69117 Heidelberg, Germany}
\affiliation{Department of Physics and Astronomy, University of Utah, Salt Lake City, UT 84112, USA}
\author[0000-0003-0248-5470]{A. C. Seth}
\affiliation{Department of Physics and Astronomy, University of Utah, Salt Lake City, UT 84112, USA}
\author[0000-0002-7489-5244]{P. J. Smith}
\affiliation{Max Planck Institute for Astronomy, K\"onigstuhl 17, D-69117 Heidelberg, Germany}
\author[0000-0001-6604-0505]{S. Kamann}
\affiliation{Astrophysics Research Institute, Liverpool John Moores University, 146 Brownlow Hill, Liverpool L3 5RF, UK}
\author[0000-0002-1670-0808]{R. Pechetti}
\affiliation{Astrophysics Research Institute, Liverpool John Moores University, 146 Brownlow Hill, Liverpool L3 5RF, UK}
\author[0000-0002-2941-4480]{M. S. Nitschai}
\affiliation{Max Planck Institute for Astronomy, K\"onigstuhl 17, D-69117 Heidelberg, Germany}
\author[0000-0002-1212-2844]{M. Alfaro-Cuello}
\affiliation{Centro de Investigación en Ciencias del Espacio y Física Teórica, Universidad Central de Chile, La Serena 1710164, Chile}
\author[0000-0002-1959-6946]{H. Baumgardt}
\affiliation{School of Mathematics and Physics, The University of Queensland, St.
Lucia, 4072, QLD, Australia}
\author[0000-0003-3858-637X]{A. Bellini}
\affiliation{Space Telescope Science Institute, 3700 San Martin Drive, Baltimore, MD 21218, USA}
\author[0000-0002-0160-7221]{A. Feldmeier-Krause}
\affiliation{Department of Astrophysics, University of Vienna, T\"urkenschanzstrasse 17, 1180 Wien, Austria}
\author[0000-0002-6072-6669]{N. Kacharov}
\affiliation{Leibniz Institute for Astrophysics, An der Sternwarte 16, 14482 Potsdam, Germany}
\author[0000-0001-9673-7397]{M. Libralato}
\affiliation{INAF, Osservatorio Astronomico di Padova, Vicolo dell’Osservatorio 5, Padova,I-35122, Italy}

\author[0000-0001-7506-930X]{A. P. Milone}
\affiliation{Dipartimento di Fisica e Astronomia “Galileo Galilei,” Univ. di Padova, Vicolo dell’Osservatorio 3, Padova, I-35122, Italy}

\author[0000-0001-8052-969X]{S. O. Souza}
\affiliation{Max Planck Institute for Astronomy, K\"onigstuhl 17, D-69117 Heidelberg, Germany}

\author[0000-0003-4546-7731]{G. van de Ven}
\affiliation{Department of Astrophysics, University of Vienna, T\"urkenschanzstrasse 17, 1180 Wien, Austria}

\author[0000-0003-2512-6892]{Z. Wang (王梓先)}
\affiliation{Department of Physics and Astronomy, University of Utah, Salt Lake City, UT 84112, USA}



\begin{abstract}

Omega Centauri (\omc) is the Milky Way's most massive globular cluster and is likely the stripped nucleus of an accreted dwarf galaxy. In this paper, we analyze \omc's kinematics using data from oMEGACat, a comprehensive catalog of \omc's central regions, including 1.4 million proper motion measurements and 300,000 spectroscopic radial velocities.  Our velocity dispersion profiles and kinematic maps are consistent with previous work but improve on their resolution, precision, and spatial coverage.  The cluster's 3D dispersion is isotropic in the core, with increasing radial anisotropy at larger radii. The 2D kinematic maps show an elongation of the velocity dispersion field comparable to the flattening observed photometrically. 
We find good agreement between proper motions and line-of-sight velocity dispersion and measure a kinematic distance of $5494\pm61$\,pc, the most precise kinematic distance to \omc{} available.
The subset of data with precise metallicity measurements shows no correlation between metallicity and kinematics, supporting the picture of well-mixed stellar populations within the half-light radius of \omc.
Finally, we study the degree of energy equipartition using a large range of stellar masses. We find partial energy equipartition in the center that decreases towards large radii. The spatial dependence of the radial energy equipartition is stronger than the tangential energy equipartition.
Our kinematic observations can serve as a new reference for future dynamical modeling efforts that will help to further disentangle the complex mass distribution within \omc.

\end{abstract}



\section{Introduction} \label{sec:intro}
\subsection{Introducing Omega Centauri}

Omega Centauri (\omc{}, NGC\,5139) is the most massive ($M\approx 3.55\times10^6$~M$_\odot$, \citealp{2018MNRAS.478.1520B}) globular cluster of our Milky Way. The stellar populations within \omc{} are complex and include an unusually wide spread in age \citep{2004A&A...422L...9H,2007ApJ...663..296V,2013ApJ...762...36J,2014ApJ...791..107V, 2016MNRAS.457.4525T,2024ApJ...977...14C} and metallicity \citep{1975ApJ...201L..71F, 2010ApJ...722.1373J,2011ApJ...731...64M, 2024ApJ...970..152N}. These complexities are also apparent in the color-magnitude diagram, which shows a multitude of different splits and sequences \citep{1997PhDT.........8A,2000ApJ...534L..83P,2004ApJ...605L.125B,2004ApJ...603L..81F,2010AJ....140..631B,2017ApJ...844..164B,2017MNRAS.464.3636M,2024arXiv241209783C}. For these reasons, \omc{} is now widely accepted to be the stripped nucleus of a dwarf galaxy that has been accreted and disrupted by the Milky Way (e.g. \citealp{1999Natur.402...55L,2003MNRAS.346L..11B}). Other evidence for this accretion scenario has been found by associating \omc{} with stellar streams in the Milky Way Halo  \citep{2012ApJ...747L..37M,2019NatAs...3..667I} and by finding potential connections with either the Sequoia or the Gaia-Enceladus merger events \citep{2019MNRAS.488.1235M,2019A&A...630L...4M,2020MNRAS.493..847F,2021MNRAS.500.2514P,2022ApJ...935..109L,2025A&A...693A.155P}.

This makes \omc{} the closest nuclear star cluster and an important witness to the formation history of the Milky Way.

Besides its peculiar stellar populations and its likely accreted origin, the internal kinematics of \omc{} have also intrigued astronomers for many years as a way to understand its mass distribution and its formation history. Traditionally, the stellar motions in \omc{} have been studied using line-of-sight velocities, limiting the observable sample to a relatively small number of a few hundred bright, evolved stars \citep{1996AJ....111.1913S, 1997AJ....114.1087M, 2006A&A...445..503R}. Early ground-based proper motion studies \citep{2000A&A...360..472V} were similarly limited to bright stars, although thousands of individual proper motions could already be measured.

The number of stars for which kinematic measurements are available has changed dramatically with the availability of multi-epoch Hubble Space Telescope (HST) data that has enabled the measurement of proper motions for hundreds of thousands of stars \citep{2010ApJ...710.1032A,2014ApJ...797..115B,2018ApJ...853...86B} down to very faint main sequence stars. More recently, the MUSE integral field spectrograph \citep{museii} at the ESO Very Large Telescope has been used to obtain spectra for hundreds of thousands of stars \citep{2018MNRAS.473.5591K, 2023ApJ...958....8N, 2024MNRAS.528.4941P}.

Previous kinematic studies of \omc{} have focused on various aspects of its kinematics including its velocity dispersion \citep{2015ApJ...803...29W}, kinematic distance \citep{2006A&A...445..513V,2015ApJ...812..149W, 2021MNRAS.505.5957B}, rotation \citep{1986A&A...166..122M,1997AJ....114.1074M,2018MNRAS.473.5591K,2024MNRAS.528.4941P,2024ApJ...970..192H}, and the energy equipartition both in the center \citep{2022ApJ...936..154W} and at larger radii \citep{2018ApJ...853...86B}. These studies showed that \omc{} is rotating with relatively high $\frac{v}{\sigma}$ leading to significant flattening. In addition, the stellar motions show partial energy equipartition and increasing radial anisotropy at larger radii.

The kinematic measurements have also served as the basis for various dynamical modeling efforts using numerous techniques to constrain the mass distribution in \omc{} which has proved to be a very complex and sometimes inconclusive task. Based on modeling of the inner region, there has been a long debate about the presence of a central, intermediate-mass black hole \citep{2008ApJ...676.1008N,2010ApJ...719L..60N,2010ApJ...710.1032A,2010ApJ...710.1063V,2017MNRAS.468.4429Z,2019MNRAS.482.4713Z,2019MNRAS.488.5340B,2025A&A...693A.104B}. The recent discovery of several high-proper motion stars near \omc{}'s center \citep{2024Natur.631..285H} provides the latest piece in this puzzle and was used to estimate a lower limit for the mass of an intermediate-mass black hole of $M_{\rm IMBH}>8200$\,M$_\odot$.

\subsection{The oMEGACat Project}
In the oMEGACat project, we have created the most comprehensive spectroscopic and astro-photometric data set for \omc{} to date. The basis for this project is two large data sets that cover the half-light radius ($r_{\rm HL}=287\arcsec$, \citealt{2018MNRAS.478.1520B}) of \omc: first, an extensive mosaic with VLT MUSE integral field observations. Based on these observations, \cite{2023ApJ...958....8N}, hereafter \citetalias{2023ApJ...958....8N}, provided a spectroscopic catalog with metallicities and line-of-sight (LOS) velocities for over 300,000 stars within the half-light radius of \omc{}. The second component of the project is a large astrometric and photometric catalog (see \citealt{2024ApJ...970..192H}, hereafter \citetalias{2024ApJ...970..192H}), which includes high-precision proper motions and multi-band photometry for around 1.4 million sources based on hundreds of new and archival HST observations.

The combined data set enables a broad range of science cases, including studies of the metallicity distribution of various subpopulations \citep{2024ApJ...970..152N}, the age-metallicity relation \citep{2024ApJ...977...14C}, the abundances of Helium \citep{2024arXiv241209783C} and other individual elements.

\subsection{This work: Overall kinematics of \omc{} in 3D}
In this work we revisit several of the key kinematic properties of \omc{} using the new combined oMEGACat catalogs, significantly extending the spatial coverage, precision, and depth of existing kinematic studies. We describe the data selection in \autoref{sec:selections}. \autoref{sec:vdisp} describes the determination of velocity dispersion, anisotropy and rotation profiles using all three velocity dimensions and the derivation of a new kinematic distance estimate. \autoref{sec:2dmaps} describes the creation of 2-dimensional kinematic maps. In \autoref{sec:metallicity} we search for potential variations of the kinematics with metallicity and in \autoref{sec:ee} we provide new detailed measurements of the state of energy equipartition. Finally, \autoref{sec:conclusions} contains a summary and conclusions.\\
Our paper stops short of studying the kinematic differences between different subpopulations and fitting dynamical models, both of which will be the content of future work. We make all products of this analysis available in electronic form to facilitate future modeling efforts. The data products released with this paper are described in \autoref{sec:dataproducts}.

\section{Data and Quality Selections}
\label{sec:selections}
The spectroscopic catalog and its creation are described in detail in \citetalias{2023ApJ...958....8N}; the HST-based astro-photometric catalog is described in \citetalias{2024ApJ...970..192H}. Here we only give a brief overview of the catalog content and describe the various quality selections used to restrict the data set to a reliable sub-sample of cluster member stars.

The spectroscopic catalog is based on observations with the VLT MUSE integral field spectrograph \citep{museii} with a total of 103 pointings. The observations were obtained for ``The MUSE Survey of Galactic Globular Clusters" (PI: S. Dreizler, S. Kamann, see also \citealt{2018MNRAS.473.5591K}) and for GO Program 105.20CG.001 (PI: N. Neumayer). Both data sets were partially observed with and without adaptive optics mode. 

The astro-photometric catalog is based on around 800 individual exposures taken with the two Hubble Space Telescope instruments ACS/WFC and WFC3/UVIS and in various different filters. The data were taken from the Archive or the dedicated Program GO-16777 (PI: A. Seth). The complete underlying data set has been collected under the following DOI: \dataset[10.17909/26qj-g090]{http://dx.doi.org/10.17909/26qj-g090}.

We photometrically reduced the data using the KS2 software \citep[see e.g.][]{2017ApJ...842....6B} and measured relative proper motions using the technique introduced in \cite{2014ApJ...797..115B}. The typical temporal baseline of the proper motion measurements is around 20.6\,yrs leading to high proper motion precision. The final catalog contains 1,395,781 sources with a proper motion measurement. On the faint end, the catalog reaches $m_{\rm F625W}\approx25$, while stars brighter than $m_{\rm F625W}<13.9$ are typically saturated.

\subsection{Selections within the HST catalog}
\subsubsection{Astrometric and photometric quality selections}
When studying the velocity dispersion, it is important to restrict the data to measurements with reliable proper motions and errors. This is especially true for energy equipartition studies, where a large range of stellar masses and, therefore, magnitudes has to be probed. In this study, we use the corrected proper motions from \citetalias{2024ApJ...970..192H}. These proper motions have been corrected for residual spatial and magnitude-dependent effects; the assumed proper motion errors are the quadratic sum of the errors of the linear proper motion fit (determined using the actual residuals) and the statistical error on the empirical correction.

We start the selection process with several global cuts on some properties of the astrometric measurements. For all stars we require the following conditions to be met (see also \autoref{fig:selectioncs_global}, Appendix, for histograms of these selections):
\begin{itemize}
    \item Temporal Baseline longer than 10 years
    \item $N_{\rm used} / N_{\rm found}$ fraction $>$ 0.75\\This parameter gives the ratio of data points that were used for the proper motion fit with respect to the total number of available measurements. A low value indicates that many data points were removed during the clipping stage, indicating unreliable astrometry.
    \item Reduced Chi-Square $<$ 5 for the proper motions fits in both components
\end{itemize}
In addition to these astrometric criteria, we require reliable photometry in the two reddest broad-band filters in the data set: ACS/WFC F625W and WFC3/UVIS F814W. The reason for these photometric quality cuts is two-fold: First, reliable photometry is needed to assess the cluster membership using color-magnitude diagrams and to estimate the mass of individual stars via isochrone fitting. Second, accurate photometry also indicates good astrometric quality, and by using two filters for which the typical time baseline is long (F625W: 2002, F814W: 2022) we can ensure that the astrometric measurements are of good quality throughout the whole monitored temporal baseline, leading to reliable proper motions.

\begin{figure*}
  \centering
    \includegraphics[width=0.9\textwidth]{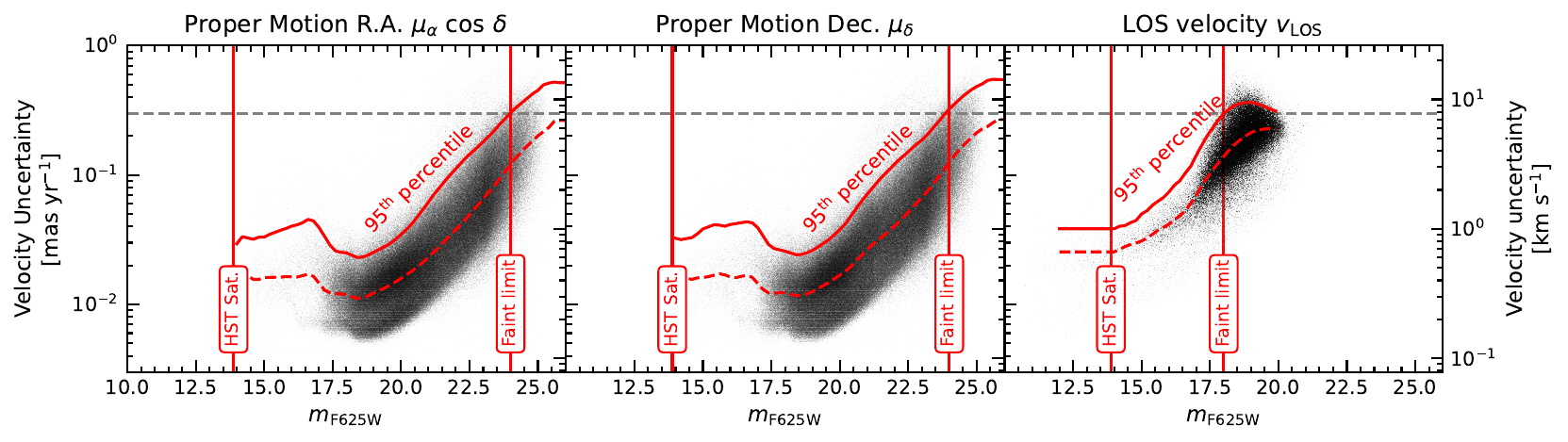}
  \caption{The three panels show the uncertainty of the individual velocity measurements for both the two proper motion directions and the line-of-sight direction plotted against the magnitude in the $m_{\rm F625W}$. The red line marks the 95th percentile of the error distribution determined in 0.5 mag wide bins and is used to reject stars with unusually large proper motion errors. The dashed line marks the median of the error distribution. To facilitate comparisons between the proper motion and the line-of-sight uncertainties all three panels have the same y scale, the left axis shows astrometric units, the right axis shows physical units at an assumed distance of 5494\,pc (this leads to a conversion of 1\,mas\,yr$^{-1}=$26.06\,km\,s$^{-1}$).}
  \label{fig:selection_pme_and_vpd}
\end{figure*}

Our photometric selections are similar to the exemplary ones described in \citetalias{2024ApJ...970..192H} and provided in the catalog, but slightly stricter. For both filters (F625W, F814W) we require:
\begin{itemize}
    \item No saturation (this leads to the exclusion of all red-giant branch stars with $m_{\rm F625W} < 13.8$)
    \item A quality-of-fit (QFIT) value higher than the 85th percentile of 0.5 mag wide intervals (using $m_{\rm F625W}$). Stars with a QFIT higher than 0.99 are always included, and stars with a QFIT lower than 0.9 are always excluded. The QFIT parameter describes how well the used point-spread-function model describes the flux distribution of each source. A value close to 1 indicates good agreement.
    \item A ratio of flux from neighboring stars within the fit aperture over the flux of the star smaller than 0.5
\end{itemize}
The combined cuts above already provide us with a reliable subsample. To ensure high and consistent quality throughout the whole magnitude range, we add one last, magnitude-dependent criterion: For both proper motion components, we require the proper motion error to be within the lower 95\% of the error distribution in 0.5~mag wide intervals (see \autoref{fig:selection_pme_and_vpd}, left). As can be seen in \autoref{fig:selection_pme_and_vpd}, this selection tracks the magnitude dependence of the bulk of all well-measured stars, while excluding outliers with unusually high errors. One can also see that at a magnitude of $m_{\rm F625W}=24$ the upper limit on the errors reaches an order of magnitude of 0.3\,mas\,yr$^{-1}$ ($\sim$7.8\,km\,s$^{-1}$). As this is similar to half of the typical velocity dispersion in the outer regions of our studied field, we exclude stars fainter than this magnitude limit. Including stars with errors similar to the actual velocity dispersion would complicate the determination of the velocity dispersion and make it quite sensitive to the modeling of the proper motion errors.

\subsection{Spectroscopic Quality selections}
For the spectroscopic catalog we defined the following criteria to create a well-measured subset of the data:
\begin{itemize}
    \item We require that each star matches the standard quality criterion defined in \citetalias{2023ApJ...958....8N}. This combined criterion contains cuts in the quality and reliability of the spectral fit, the accuracy of the recovered magnitude, the average signal-to-noise ratio, and the cluster membership. We further restrict the kinematic subsample to measurements with a relative mag accuracy \texttt{mag\_rel}$>0.9$ and a reliability parameter \texttt{rel}$>0.9$. This helps to remove stars that are influenced by neighboring sources and may bias the kinematic measurements.
    \item Similar to the HST-based proper motions, we rejected measurements whose line-of-sight velocity error was larger than the 95th percentile in 0.5 mag wide bins (see \autoref{fig:selection_pme_and_vpd}, \textit{right}). 
    \item We set an overall magnitude cutoff at $m_{\rm F625W}>18$. At this magnitude our magnitude-dependent error cutoff reaches a level of $\sim7.7\,$km\,s$^{-1}$ (equivalent to the 0.3\,mas\,yr$^{-1}$ cutoff of $m_{\rm F625W}=24$ for the proper motion measurements).
    \item We also require a successful crossmatch with the HST-based catalog (this was achieved for 307,030 of 342,797 stars from the MUSE catalog; see \citetalias{2024ApJ...970..192H}) and a high-quality HST-based measurement. This effectively makes the MUSE sample a subset of the HST sample and allows us to apply the same membership selections. In addition, it makes the MUSE subset a true 3D sample allowing us to compare the results for both proper motion and line-of-sight based measurements. The final MUSE sample contains 32,092 stars with a high-quality LOS measurement.

\end{itemize}

\subsection{Cluster membership selection}
\begin{figure*}
  \centering
    \includegraphics[width=0.9\textwidth]{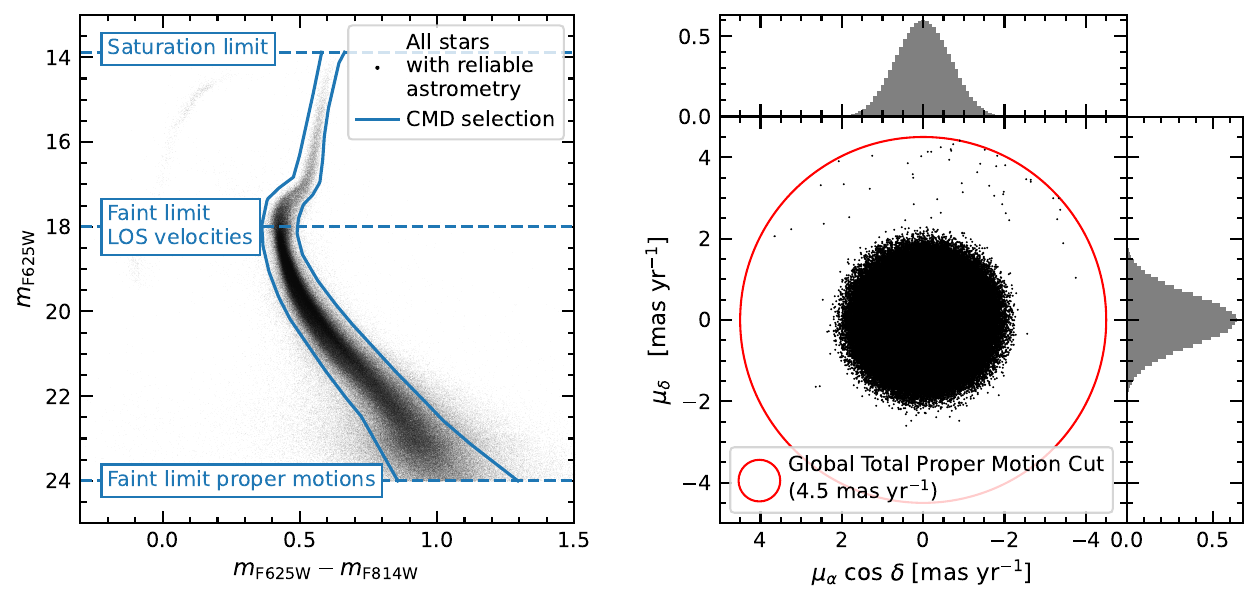}
  \caption{\textit{Left: } A color-magnitude diagram based on photometry in the $m_{\rm F625W}$ and the $m_{\rm F814W}$ filter for all stars that pass the basic quality selections for the subsample used for the kinematic analysis. The blue solid lines mark our selection of main sequence and red-giant branch stars that are members of \omc{}. The dashed lines mark the bright and faint limits of the subsample used for the kinematic analysis. \textit{Right: } A vector-point diagram of the stars in the high-quality subsample. The red circle marks the global proper motion cutoff of 4.5\,mas\,yr$^{-1}$. Stars with a total proper motion higher than this value are excluded from the kinematic analysis as they are likely fore- or background stars. The histograms in the side panels show the marginalized distributions of the two proper motion components.}
  \label{fig:selection_cmd_and_map}
\end{figure*}

To restrict our sample to likely cluster members and exclude fore- and background sources, we use both a photometric and a proper-motion-based criterion. First, we require that the stars lie on the red-giant branch or the main sequence in the F625W-F814W color-magnitude-diagram using two manually defined fiducial lines (see \autoref{fig:selection_cmd_and_map}, \textit{left}). This excludes cluster stars on the horizontal branch, however, their numbers are comparatively low and no MUSE line-of-sight velocities were measured for them. The other criterion is a global cut-off in total proper motion of 4.5\,mas\,yr$^{-1}$. This corresponds to around 115\,km\,s$^{-1}$, around 5.5 times higher than the typical velocity dispersion for main-sequence stars in the center of \omc{}. There is a small number of stars that pass this criterion but are likely non-members (see stars in upper-right of the vector-point diagram in \autoref{fig:selection_cmd_and_map}). These sources are removed with an additional sigma-clipping step when determining the actual kinematic properties.

\subsection{Summary of selections}
From an initial number of 1,395,781 stars with a proper motion measurement, 669,975 pass our combined quality selection criteria, of which \underline{610,846} then pass the subsequent membership cuts, constituting our proper motion sample.

\begin{figure}[h]
  \centering
    \includegraphics[width=0.45\textwidth]{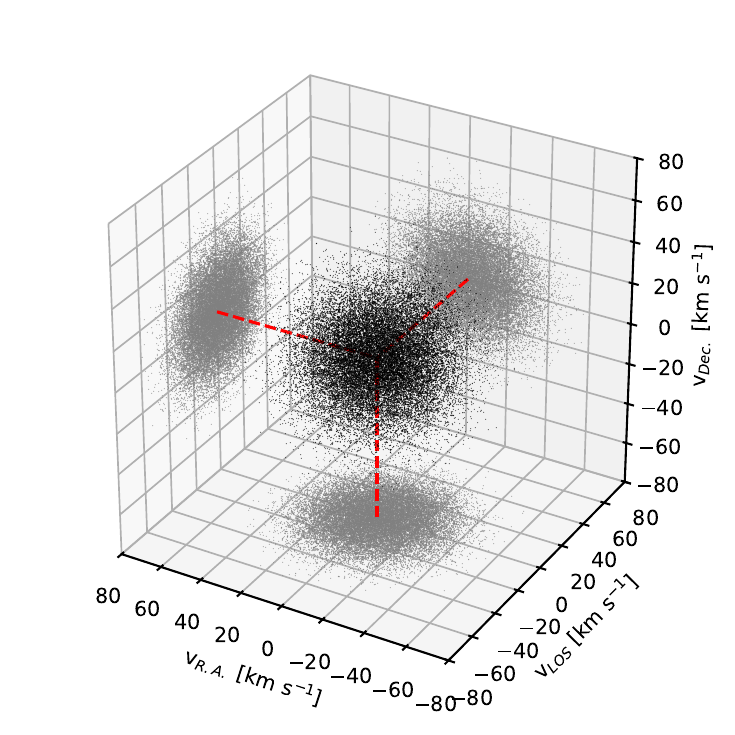}
  \caption{3D vector point diagram for the 24,928 stars that match both our proper motion and line-of-sight velocity quality criteria. Proper motions have been converted to physical velocities using our new kinematic distance of $d=$5494\,pc.}
  \label{fig:vpd3d}
\end{figure}

From 342,797 stars with a line-of-sight velocity, 307,030 were successfully crossmatched with the HST catalog, of which 32,092 passed all spectroscopic quality criteria. Finally, \underline{24,928} stars have \underline{both} a high-quality astrometric and photometric measurement. The full 3D sample of velocities is shown in a 3-dimensional version of a vector point diagram in \autoref{fig:vpd3d}. This Figure shows that all 3 velocity components show a similar distribution when assuming a distance of $d=5494\,$pc (our best fit kinematic distance, see \autoref{subsec:kindistance}).

\section{1D profiles of the kinematic parameters}
\label{sec:vdisp}
\subsection{Determination of the velocity dispersion}
The observed velocity distribution for both proper motions and LOS velocity is a superposition of the true velocity distribution and the measurement errors. To measure the underlying velocity dispersion we used the log-likelihood function for a Gaussian distribution with heterogeneous errors in the form presented by \cite{1993ASPC...50..357P}:

\begin{equation}
    LL = -\frac{1}{2}\sum_i \left(\frac{(v_i - \bar{v})^2}{(\sigma^2 + \sigma_i^2)} + \text{log} (2\pi (\sigma^2 + \sigma_i^2) )\right)
\end{equation}
with $\bar{v}$ being the mean velocity in a certain subsample, $\sigma$ being the true velocity dispersion, and $v_i$ and $\sigma_{i}$ being the individual stellar velocity measurements with their uncertainties.
We sampled the likelihood function using the Markov-Chain Monte-Carlo code \texttt{emcee} \citep{2013PASP..125..306F}, using flat priors and 12 walkers with 500 steps each. We use the median of the posterior distribution as our best estimate for the velocity dispersion and use the 16th and 84th percentile as measures of the uncertainty.
\subsection{Proper Motion based 1D profiles of the velocity dispersion.}
To measure spatial variations of the velocity dispersion the data is typically split into radial bins. Several binning schemes are possible, and there is a trade-off between spatial resolution, stochastic noise, and ease of presentation. We compare several binning schemes in \autoref{app:binning} but choose an adaptive logarithmic binning scheme as our standard. The radii of successive bins are increased by a factor of at least $10^{0.05}\approx1.122$, while maintaining a minimum number of 100 stars per bin.

For all radial profiles and for the decomposition of the proper motions in radial and tangential components we adopt the photometric center of the cluster as determined by \cite{2010ApJ...710.1032A} with \texttt{R.A. = 13:26:47.24h} and \texttt{Dec. = -47:28:46.45$^\circ$}. This center is independently confirmed by the presence of fast-moving stars \citep{2024ApJ...970..192H} and by the kinematic center estimate performed in this work (see \autoref{subsec:gaussianfits}).

\begin{figure*}[h]
  \centering
    \includegraphics[width=0.9\textwidth]{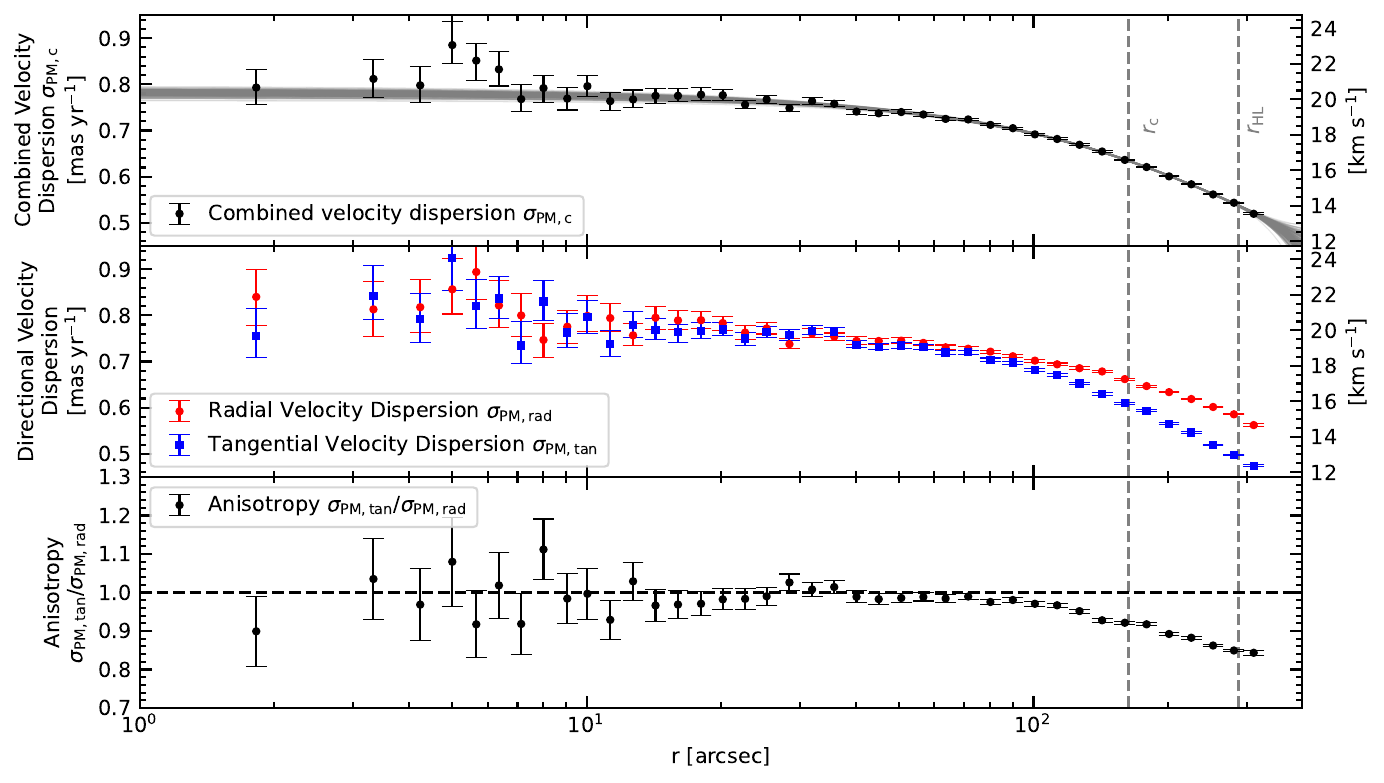}
  \caption{Proper motion dispersion profiles determined using the new oMEGACat data using the full magnitude range. The profiles were determined using an adaptive logarithmic binning scheme with a stepsize of $\Delta \text{log} r = 0.05$ and a minimum number of 100 stars per bin. The dashed vertical lines indicate the core and the half-light radii as reported in \cite{2018MNRAS.478.1520B} \textit{Top:} Overall dispersion, for which measurements of the tangential and the radial component of the proper motion were combined. The grey line shows the result of 100 4th-order polynomial fits to the dispersion profile and is meant for visualization purposes only. \textit{Center:} Individual components of the proper motion dispersion in which the tangential and the radial components were treated separately. \textit{Bottom:} Anisotropy profile calculated as the ratio between the tangential and the radial proper motion dispersion component.}
  \label{fig:1d_profiles}
\end{figure*}

For the overall (combined) proper motion dispersion $\sigma_{\rm PM,c}$ we treat the radial and tangential components of the proper motion as separate samples from the velocity distribution, which doubles the number of measurements. The resulting profile is shown in \autoref{fig:1d_profiles}, and the individual numerical values can be found in \autoref{tab:dispersionpm} (\autoref{sec:dataproducts}). The velocity dispersion rises steadily from 0.52\,mas\,yr$^{-1}$ (13.6\,km\,s$^{-1}$) at large radii to the central 10 arcseconds, where it reaches a mean value of $\sim$0.81\,mas\,yr$^{-1}$ (21.1\,km\,s$^{-1}$).  The error bars at large radii are as small as 0.001\,mas\,yr$^{-1}$ but are higher near the center due to the smaller number of stars per bin.

\subsubsection{Comparison with literature profiles}
\begin{figure}[h]
  \centering
    \includegraphics[width=0.45\textwidth]{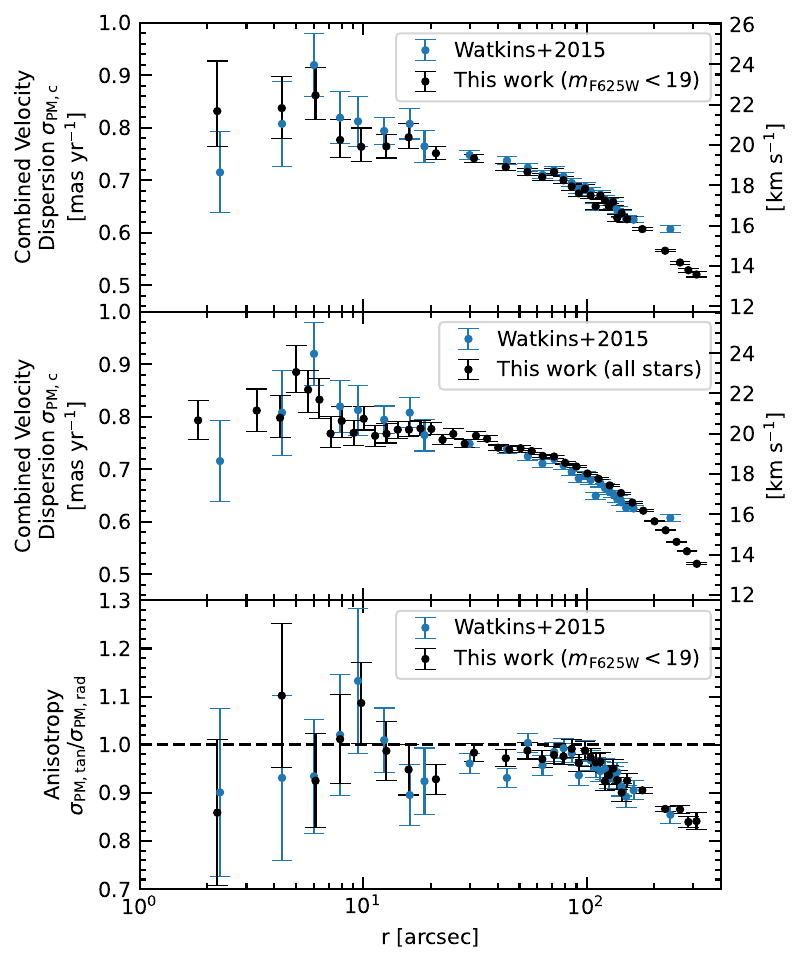}
  \caption{Proper motion dispersion profiles determined using the new oMEGACat (black markers). We compare the new dispersion measurements with the literature profile by \citetalias{2015ApJ...803...29W} (blue markers). For better comparability, in the \textit{top panel} we restrict our data set to bright stars and use a binning scheme similar to the literature. In the \textit{center panel} we use logarithmic radial bins and the full high-quality subset. In the \textit{bottom panel} we compare the anisotropy, again using the bright sample only.}
  \label{fig:watkins_comparison}
\end{figure}

The previous most widely used profile of the proper motion dispersion in the inner region of \omc{} has been published in \cite{2015ApJ...803...29W} (thereafter \citetalias{2015ApJ...803...29W}). In \autoref{fig:watkins_comparison} we compare the literature profile with our new profile. For the comparison, we have to take into account that the \citetalias{2015ApJ...803...29W} profile is based on a subset of bright stars. Due to the partial energy equipartition in the core of \omc{}, we expect a higher dispersion measured from our catalog, as we include lower mass stars in the analysis. To allow for a consistent comparison, we also calculated a profile using only stars brighter than $m_{\rm F625W}=19$, a threshold similar to the one used in \citetalias{2015ApJ...803...29W} and using a similar binning scheme.

The comparison of the profiles (\autoref{fig:watkins_comparison}, \textit{middle panel}) matches our expectations: Due to the significantly larger number of included measurements both the spatial resolution and the individual errors in the new dispersion profile are improved when all well-measured stars from the new proper motion catalog are used. Our profile shows less scatter and extends to larger radii. At larger radii, it shows slightly larger dispersion values as expected from energy equipartition arguments: Using the same technique as in \autoref{sec:ee}, we derive a mean stellar mass of 0.675\,M$_\odot$ within the \citetalias{2015ApJ...803...29W} sample and of 0.515\,M$_\odot$ within our full sample (that is spanning all magnitudes). Using the central energy equipartition value of $\eta=0.08$ (\autoref{sec:ee}) and \autoref{eq:eeeta} predicts a 2\% larger velocity dispersion when using the full sample.

When comparing a similar sample of bright stars (\autoref{fig:watkins_comparison}, \textit{top panel}) we see similar errors in the dispersion measurements (as these errors are dominated by the limited number of available stars in each bin and not on the individual proper motion measurement errors) and overall agreement between the two profiles. However, we notice an overall smaller dispersion in all but the two innermost bins. One potential explanation could be a small underestimation of the proper motion errors in the literature work, leading to an overestimation of the velocity dispersion.

\subsection{Velocity Anisotropy Profiles}
To study the velocity dispersion anisotropy, we decompose the proper motion measurements into their radial and tangential components (with respect to the \citealt{2010ApJ...710.1032A} cluster center). As there are no strong correlations between the measurements in the right ascension and declination components, we can treat them as independent measurements and calculate the errors on the projected components accordingly. We then calculate the dispersion profile for the two components separately (\autoref{fig:1d_profiles}, middle panel). While there are no apparent differences in the central regions, at larger radii, the radial velocity dispersion ($\sigma_{\rm PM,rad}$) is significantly higher than the tangential velocity dispersion ($\sigma_{\rm PM,tan}$). To quantify this we also calculate the ratio between the two dispersion values $\sigma_{\rm PM,tan}/\sigma_{\rm PM,rad}$, see bottom panel in Fig.~\ref{fig:1d_profiles}. We find no significant anisotropy within $r<30\arcsec$; after that, the velocity distributions become increasingly radially anisotropic, reaching $\sigma_{\rm PM,tan}/\sigma_{\rm PM,rad}=0.0849\pm0.003$ at $281\arcsec$ close to the half-light radius. 

There is good agreement between our new measurements and the anisotropy profiles derived in \citetalias{2015ApJ...803...29W}, see bottom panel of \autoref{fig:watkins_comparison}. However, the new measurements reach significantly larger radii.

\subsubsection{Comparison with other clusters}

\cite{2022ApJ...934..150L} derived detailed kinematics for a large sample of Milky Way globular clusters and related the velocity dispersion anisotropy at the half-light radius with the half-light radius relaxation time (see their Figure 6). With an anisotropy value of $\sigma_{\rm PM,tan}/\sigma_{\rm PM,rad}=0.0849\pm0.003$, the half-light anisotropy in \omc{} is significantly lower than for most other Milky Way globular clusters. Due to its young dynamical age (half-mass relaxation time $\sim21$\,Gyr; \citealt{2018MNRAS.478.1520B}) it still follows the trends presented in \cite{2022ApJ...934..150L}. 

\subsection{Dispersion and rotation profiles based on the LOS Data}

Unlike the proper motions, which have been measured relative to the bulk motion of the cluster and therefore do not contain any rotation signal, the line-of-sight velocity measurements are absolute and do show the rotation. Thus, the velocity dispersion and the rotation profile have to be determined simultaneously. 
We do this using the method developed in \cite{2018MNRAS.473.5591K}, which simultaneously fits for the velocity dispersion $\sigma_{\rm LOS}$, the rotation $v_{\rm LOS}$ and the position angle of the rotation axis $\theta_0$ in each bin.

\begin{figure*}[]
  \centering
    \includegraphics[width=0.9\textwidth]{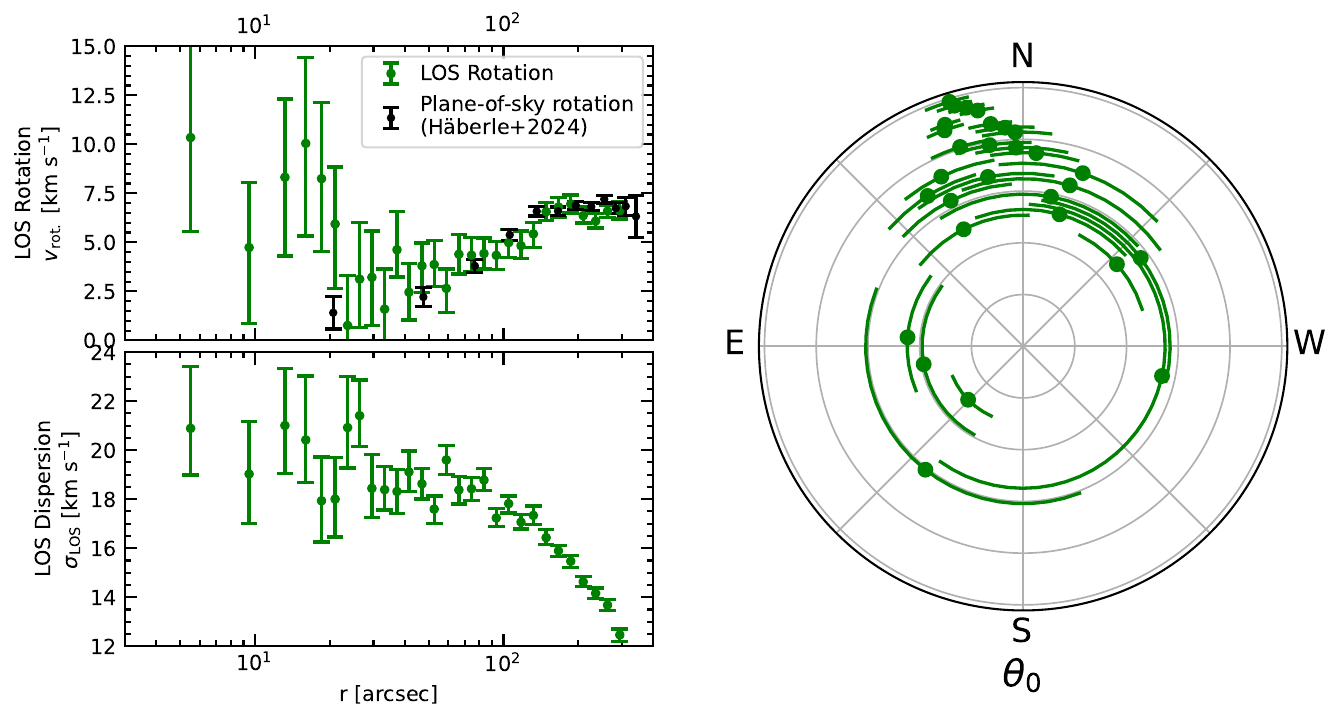}
  \caption{\textit{Top left: }Rotation profile determined using the line-of-sight velocities. We compare our new LOS rotation profile with the plane-of-sky rotation curve presented in \cite{2024ApJ...970..192H} and find good agreement. \textit{Bottom left:} Line-of-sight velocity dispersion profile. \textit{Right:} Result for the position angle of the determined rotation axis.}
  \label{fig:los_profiles}
\end{figure*}

Again we use an adaptive logarithmic binning scheme ($\Delta$log$r=0.05$, $N_{\rm min}=50$) to split the data into circular bins. The resulting dispersion and rotation curves of the LOS velocities are shown in  \autoref{fig:los_profiles} and in \autoref{tab:dispersionlos} (\autoref{sec:dataproducts}).

The measured rotation curve starts with relatively high values in the innermost bins (although with uncertainties as large as $\sim$5\,km\,s$^{-1}$ due to the small number of measurements) before reaching a minimum at around $r=$30\arcsec. Afterwards it increases monotonically until reaching a plateau with $v_{\rm rot }=7$\,km\,s$^{-1}$ at around $r=$150\arcsec. The mean value of the position angle of the rotation axis for $r>30\arcsec$ is $\theta_{\rm LOS} = (104.3\pm1.4)^\circ$.

The initial decrease of the rotational velocity is likely related to the counter-rotating structure discovered in \cite{2024MNRAS.528.4941P} as also indicated by the flip of the rotation angle (see \autoref{fig:los_profiles}, \textit{right}). The constant rotation velocity at radii larger than $r=$150\arcsec\ is also observed in the plane-of-sky rotation (see \citealt{2024ApJ...970..192H}), as shown by the comparison in \autoref{fig:los_profiles}.  The similar rotation amplitudes of the different spatial components are expected due to the inclination of $i=43\fdg9\pm1\fdg3$ as determined in \cite{2024ApJ...970..192H}.

\subsection{Comparison between proper-motion and line-of-sight dispersion profiles and kinematic distance}
\label{subsec:kindistance}
As the kinematic LOS sample is a subset of the proper motion sample, we can also calculate the proper motion profile using the same stars for all dimensions. 
This allows us to determine the kinematic distance of \omc{}, but also to capture potential systematic effects in either data set.

We use the following equation to obtain a kinematic distance estimate for each individual bin of the line-of-sight velocity dispersion profile, assuming that the proper motion dispersion and the line-of-sight velocity dispersion are the same:
\begin{equation}
    d\,[{\rm pc}] = (210.51\,[{\rm pc\,km^{-1}\,s\,yr^{-1}}])\cdot\frac{\sigma_{\rm LOS}\,[{\rm km\,s}^{-1}]}{\sigma_{\rm PM, c}\,[{\rm mas\,yr}^{-1}]}
\end{equation}

\begin{figure}[h]
  \centering
    \includegraphics[width=0.45\textwidth]{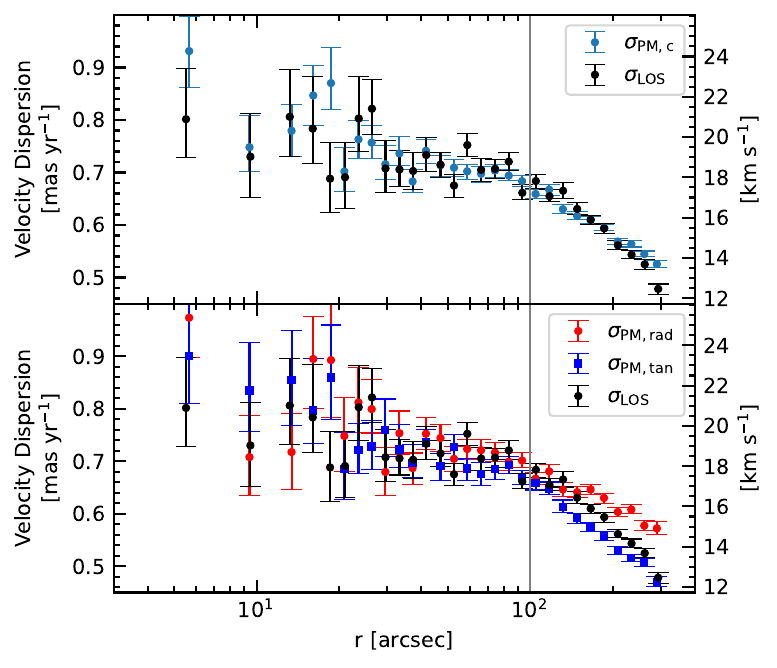}
  \caption{\textit{Top: } A comparison between the line-of-sight velocity dispersion profile and the overall proper motion dispersion. A distance of 5494\,pc was assumed to convert between proper motions and physical velocities. The two profiles show good agreement. \textit{Bottom: } A comparison between the line-of-sight velocity dispersion and the individual (radial and tangential) components of the proper motion dispersion. At larger radii ($r>100\arcsec$, gray vertical line), where the proper motion field turns increasingly radially anisotropic, the line-of-sight velocity dispersion lies in between the two different proper motion profiles.}
  \label{fig:los_profile_comparison}
\end{figure}

The upper panel of \autoref{fig:los_profile_comparison} shows a comparison between the proper motion dispersion ($\sigma_{\rm PM,c}$) and the line-of-sight velocity dispersion ($\sigma_{\rm LOS}$) using a distance of 5494\,pc, the best fit distance described below. The two profiles show good agreement within their error bars for most of the bins. In the lower panel of \autoref{fig:los_profile_comparison} we compare the line-of-sight profile with the radial and tangential components of the proper motion dispersion. At larger radii, where the velocity anisotropy is more pronounced, one can see that the line-of-sight dispersion falls in between the radial and tangential proper motion dispersion profile. This can be explained geometrically: the LOS velocities contain both a radial and a tangential component, depending on the (unknown) LOS position of the star.

\begin{figure}
  \centering
    \includegraphics[width=0.45\textwidth]{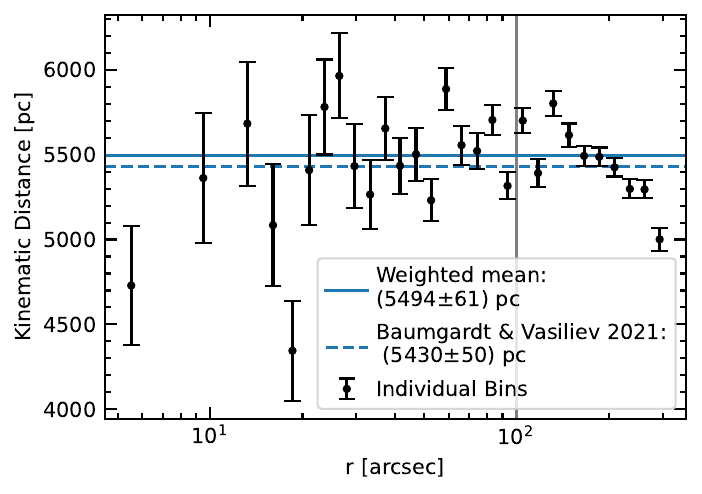}
  \caption{Kinematic distance of \omc{} derived by calculating the ratio between line-of-sight and proper motion dispersion in different circular bins (see also \autoref{fig:los_profile_comparison}, \textit{top}). We restrict the calculation of a weighted mean velocity to the region $r<100\arcsec$ (gray vertical line), to limit the influence of anisotropy.} The weighted mean value is in good agreement with the literature distance value derived in \cite{2021MNRAS.505.5957B}.
  \label{fig:kinematic_distance}
\end{figure}

The results for the kinematic distance are shown in \autoref{fig:kinematic_distance}.  To limit the influence of the anisotropy on the kinematic distance estimate, we restrict our analysis to the inner region of \omc{}  ($r<100\arcsec$), for which the velocity dispersion is approximately isotropic (see \autoref{fig:1d_profiles}). The variance weighted mean of all individual kinematic distance estimates within our cutoff radius is ($5494\pm61$)\,pc. This value is in 1$\sigma$ agreement with the value of (5430$\pm$50)\,pc determined in \cite{2021MNRAS.505.5957B} by averaging several different distance estimation methods (see their paper for a detailed comparison of various other literature distance estimates). As our estimate is based on a consistent dataset (with the same large sample of stars with both well-measured proper motions and LOS velocities) we consider it the most reliable available kinematic distance value, in addition to being one of the most precise distance measurement of any kind available for \omc. An overview of various other distance estimates including Gaia parallaxes is given in \autoref{tab:distances} in the Appendix.

We note that our simple method of estimating the kinematic distance requires the assumption of isotropy between the proper motion and the line-of-sight velocity components which is why we restrict ourselves to radii $r<100\arcsec$. In this inner region, the anisotropy is close to one ($\sigma_{\rm PM,tan}/\sigma_{\rm PM,rad}\geq0.97$). At larger radii we expect some bias due to anisotropy and flattening of the velocity field \citep[see also][]{2006A&A...445..513V}.  
The lower kinematic distance values for the bins at larger radii (see \autoref{fig:kinematic_distance}) might be caused by this effect; the weighted mean for the kinematic distance using all available bins is ($5445\pm41$)\,pc.
The Gaia kinematic distance estimate from \citet{2021MNRAS.505.5957B} of ($5359\pm141$)\,pc is derived from data at predominantly larger radii than our estimate; and is therefore independent of our value; the consistency in the derived distance therefore adds further credibility to our estimate and suggests our kinematic distance is reliable. However, modeling of the oMEGACat data based on an accurate anisotropic, rotating, and flattened model fit to the data could result in an improved estimate and will be subject of a future paper.

\clearpage
\section{Kinematic Maps}
\label{sec:2dmaps}
The large number of stellar measurements in the new oMEGACat catalogs allows us to derive kinematic maps with fine spatial resolution and large spatial coverage.
\subsection{Maps of the total proper motion dispersion}
\begin{figure*}[h]
  \centering
    \includegraphics[width=0.9\textwidth]{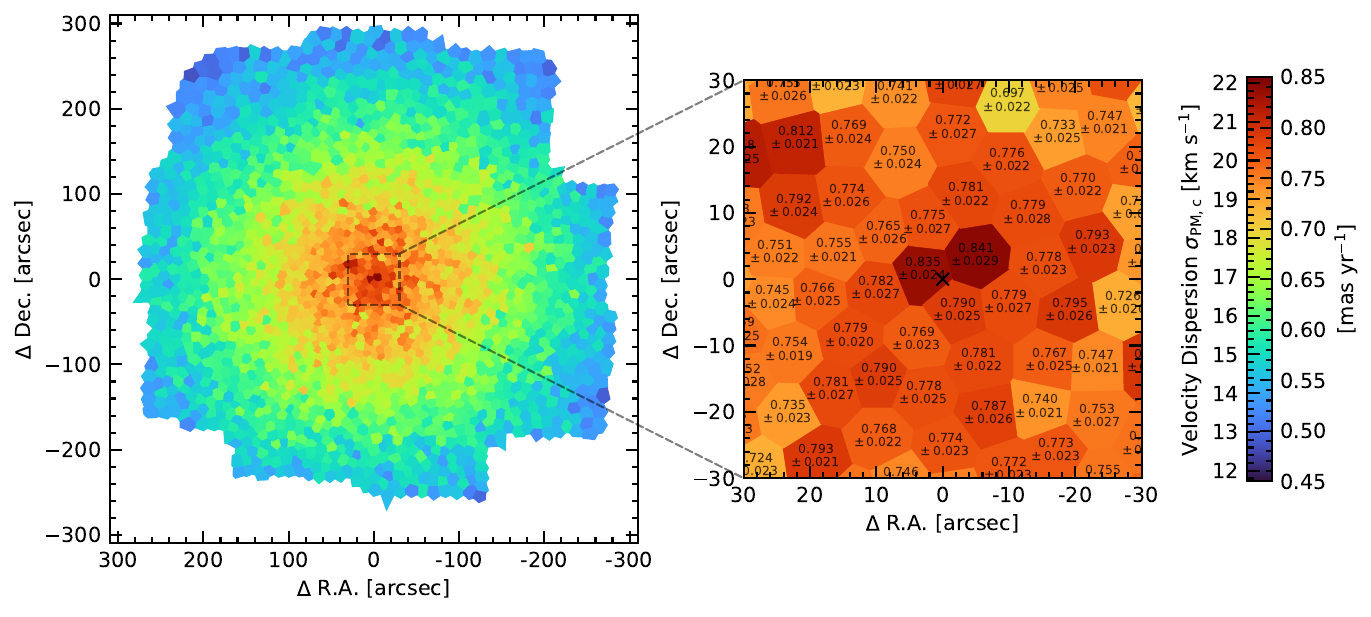}
  \caption{Dispersion map combining both proper motion components and determined using $N=250$ Voronoi bins. The right panel shows a zoom into the centermost arcminute with the numerical values (in mas\,yr$^{-1}$) for the individual bins shown in black letters.}
  \label{fig:voronoi_map_proper_motion}
\end{figure*}

To create velocity dispersion maps of the proper motion we use the Python Voronoi binning package \texttt{vorbin} \citep{2003MNRAS.342..345C} to separate the field into approximately equally populated 2-dimensional bins. 
We set a target number of 250 stars per bin, which yields a median uncertainty of approximately 0.02\,mas\,yr$^{-1}$ ($\approx$\,0.5\,km\,s$^{-1}$) per bin and a total number of 2,434 bins. \autoref{fig:voronoi_map_proper_motion} shows the resulting map of the combined velocity dispersion $\sigma_{\rm PM,c}$ with a zoom into the innermost arcminute. The map shows the overall decrease of the velocity dispersion towards larger radii and has an overall symmetric and smooth appearance, indicative of the large number and high quality of the underlying velocity measurements.

\subsection{Empirical fits of a smooth model}
\label{subsec:gaussianfits}
\begin{figure*}[h]
  \centering
    \includegraphics[width=0.9\textwidth]{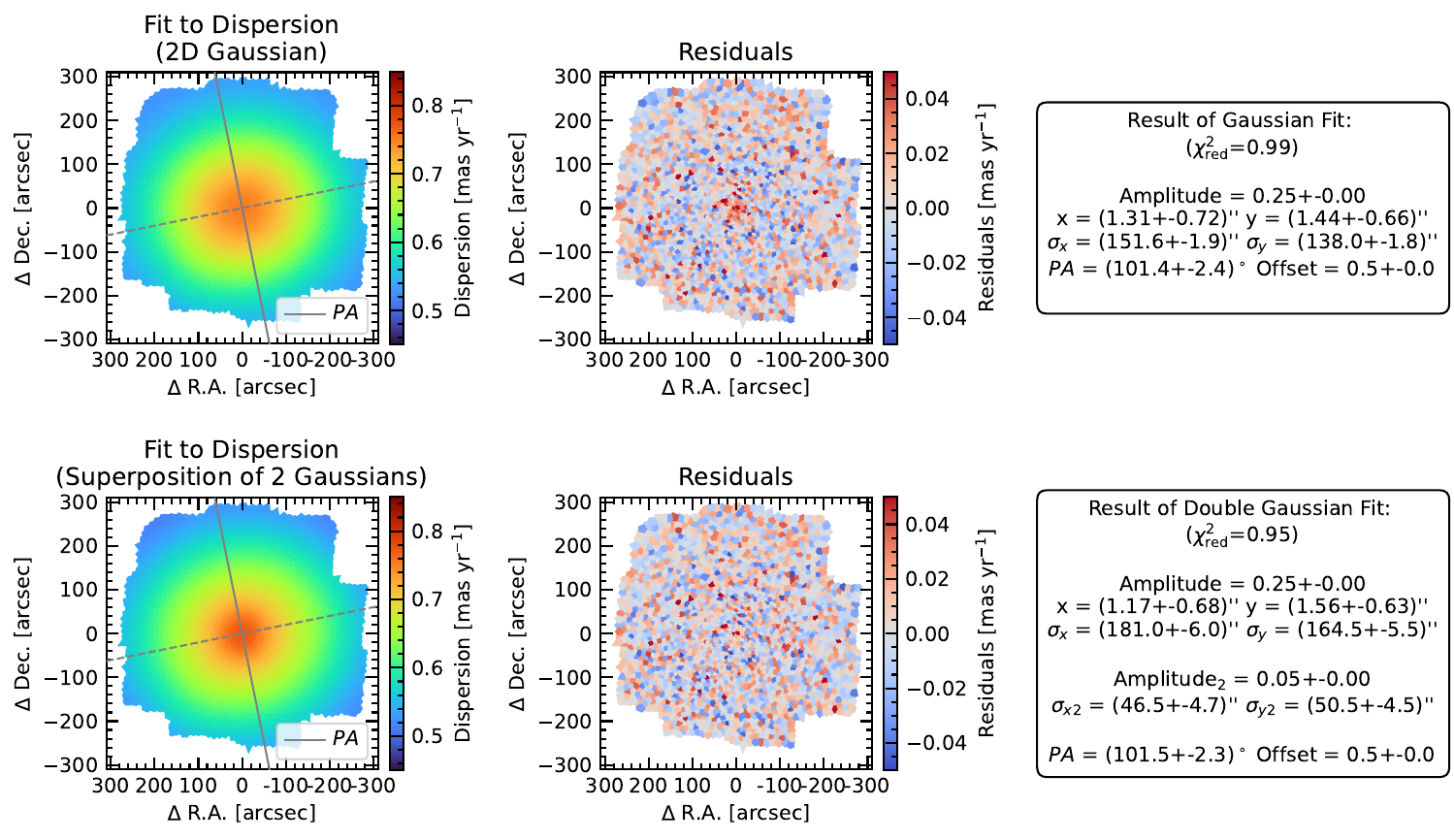}
  \caption{\textit{Top:} Result of a single component 2D Gaussian fit to the 2D proper motion dispersion field shown from \autoref{fig:voronoi_map_proper_motion}. The \textit{left} shows the result, the \textit{center} shows the residuals, which are in good agreement besides an underestimation of the cusp of the velocity dispersion in the centermost region, and the \textit{right} shows the parameters of the Gaussian fit. \textit{Bottom:} Results for a 2-component 2D Gaussian fit. This 2-component model is better able to describe the velocity dispersion in the innermost region, as can be seen from the residuals. Both models successfully recover the ellipticity and the position angle of the dispersion field.}
  \label{fig:voronoi_map_gaussian_fit}
\end{figure*}

To determine the general geometric properties of the velocity dispersion map and a kinematic estimate for the cluster center, we fit a 2D Gaussian function to velocity dispersion values determined in the Voronoi bins. A single 2D Gaussian provides a decent fit (see \autoref{fig:voronoi_map_gaussian_fit}) with a reduced $\chi^2=0.99$ and independently recovers the cluster center ($\Delta\mathrm{R.A.}=-1\farcs31\pm0\farcs72$, $\Delta\mathrm{Dec.}=1\farcs44\pm0\farcs66$) with respect to the photometric center determined in \citealt{2010ApJ...710.1032A}). The best-fit position angle is $\Theta=101\fdg4\pm2\fdg4$ (counterclockwise offset of the major axis with respect to North), and the dispersion distribution is flattened along the minor axis of the cluster with $1-\sigma_y/\sigma_x = 0.09$. The flattening of the velocity field is similar to the photometric flattening, where the mean ellipticity $\epsilon = 1-\frac{b}{a} = 0.010$ \citep{1983A&A...125..359G,2003MNRAS.345..683P,2020ApJ...891..167C}, see also \autoref{app:binning}. The position angle also is in agreement with the photometric value (e.g. 100$^\circ$, \citealt{2006A&A...445..513V}) and with the rotation axis value found using LOS velocities ($\Theta_{\rm 0,LOS}=104\fdg3\pm1\fdg4$, see above).

Even though the global properties are well described by a single Gaussian, the residuals show that there is a significant rise of the velocity dispersion within $r<10\arcsec$. Future dynamical models are necessary to interpret this further, but the size of this feature is comparable to the radius of influence of a $\sim$20,000\,M$_\odot$ intermediate-mass black hole using the equation of \cite{1972ApJ...178..371P}.
Allowing a second Gaussian component (with the same center as the first component) in the fit model (see \autoref{fig:voronoi_map_gaussian_fit}, \textit{lower row}) allows the model to describe this central rise in velocity dispersion and to further reduce the reduced $\chi^2$ to 0.91. The kinematic center, the position angle, and the flattening of the outer component are still successfully recovered.

\subsection{Maps of the proper motion anisotropy}
\begin{figure*}[h]
  \centering
    \includegraphics[width=0.9\textwidth]{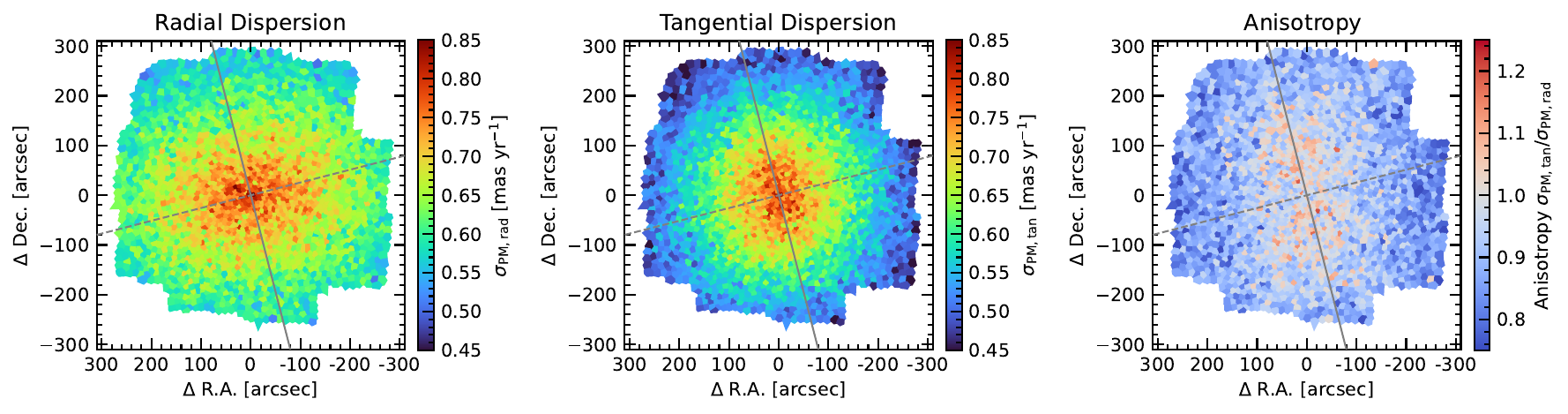}
  \caption{Proper motion dispersion maps determined separately for the radial (\textit{left}) and the tangential (\textit{center}) proper motion component. These individual-component velocity fields show strong flattening/elongation along the rotation axis (solid grey line) of the cluster, which is likely a geometric effect. The ratio of the two components (\textit{right}) gives a measure of the anisotropy of the velocity fields. At larger radii, the field becomes increasingly radially anisotropic. In addition, there are two tangentially anisotropic plumes along the rotation axis.}
  \label{fig:anisotropy_maps}
\end{figure*}

To study the 2-dimensional variation of the velocity anisotropy we use the same binning scheme as in the previous section, but now calculate the two dispersion components ($\sigma_{\rm PM,rad}$ and $\sigma_{\rm PM,tan}$) separately, see \autoref{fig:anisotropy_maps}. While the overall velocity dispersion distribution shows only mild ellipticity, the radial velocity dispersion appears highly flattened with respect to the rotation axis of the cluster and the tangential velocity dispersion appears to be elongated along the rotation axis. The 2D map of the anisotropy ($\sigma_{\rm tan}/\sigma_{\rm rad}$) shows the overall trend to radial anisotropy at larger radii, but also two ``tangentially anisotropic" plumes along the rotation axis. This 2D structure can be naturally explained as a superposition of the actual physical anisotropy in the velocity dispersion and a geometric projection effect of the rotation of the cluster: In a 2D projection, stars that are close to the rotation axis preferentially move orthogonally to the axis resulting in an apparent increase in the observed tangential velocity dispersion. 
\vspace{-0.1cm}

\subsection{Maps of the line-of-sight mean velocity and dispersion}

\begin{figure*}[h]
  \centering
    \includegraphics[width=0.9\textwidth]{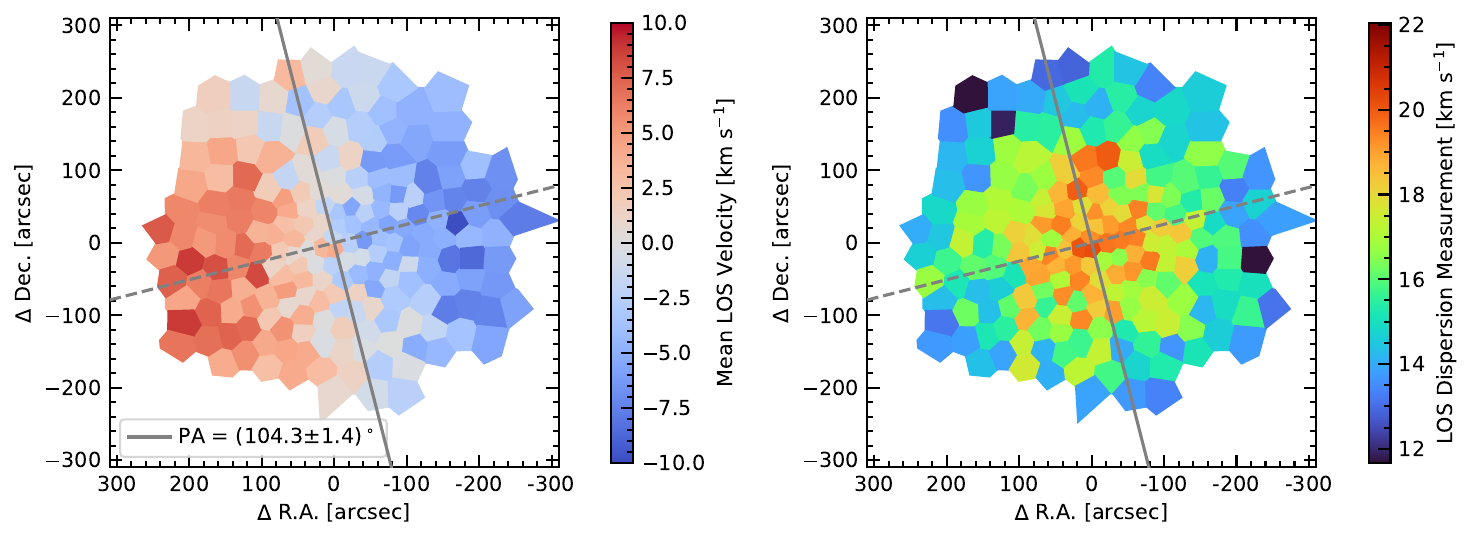}
  \caption{\textit{Left: } Mean line-of-sight velocity determined in $N=100$ Voronoi bins. The global rotation of \omc{} is clearly visible. \textit{Right: } Line-of-sight velocity dispersion determined in Voronoi bins.}
  \label{fig:los_maps}
\end{figure*}

In \autoref{fig:los_maps} we derive Voronoi binned kinematic maps based on the line-of-sight velocity measurements. Due to the significantly smaller size of the LOS sample (24,928 sources compared to 610,846 sources in the proper motion sample) we resorted to a smaller target bin size of $N=100$ stars per bin. This still leads to larger bins compared to \autoref{fig:voronoi_map_proper_motion}.

\begin{figure*}[h]
  \centering
    \includegraphics[width=0.9\textwidth]{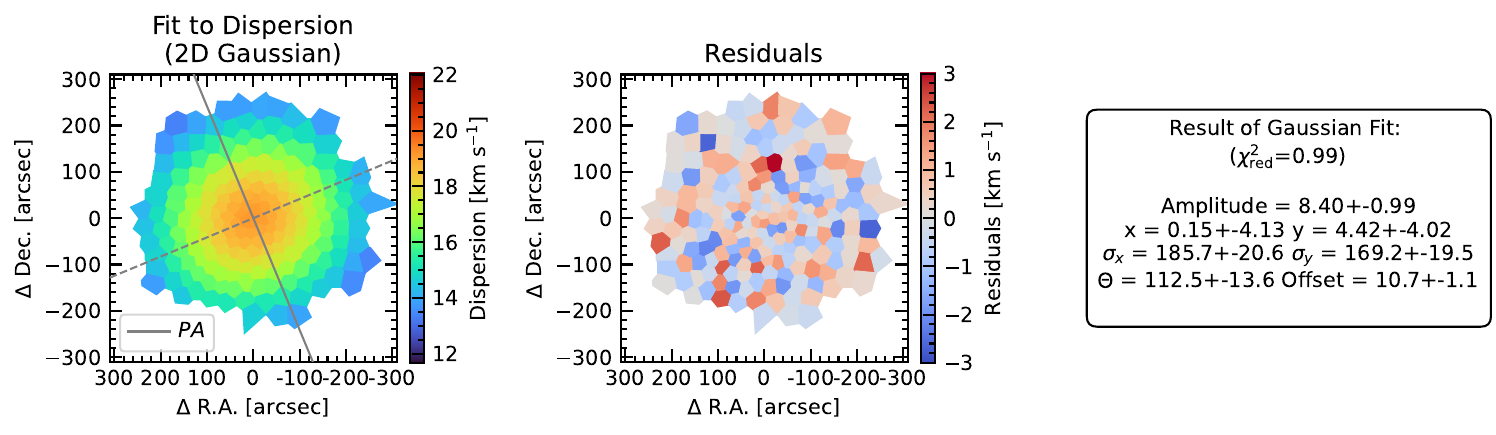}
  \caption{Result of a 2D Gaussian model fit to the line-of-sight velocity dispersion map (compare with \autoref{fig:voronoi_map_gaussian_fit}). \textit{Left:} Best fit Gaussian model \textit{Center:} Residuals \textit{Right:} Parameters of the fit.} 
  \label{fig:los_maps_with_fit}
\end{figure*}

Opposite to the proper motion measurements (which by construction have a zero mean motion) the line-of-sight velocities contain the rotation signal of the cluster. Therefore we show both the mean line-of-sight velocity per bin (\autoref{fig:los_maps}, \textit{left}) and the derived velocity dispersion $\sigma_{\rm LOS}$(\autoref{fig:los_maps}, \textit{right}). The mean velocity map nicely shows the line-of-sight rotation pattern. The dispersion map looks similar to the maps derived with the proper motions, however, it shows larger scatter (as expected due to the smaller sample size). We performed a 2D Gaussian fit to the dispersion field (\autoref{fig:los_maps_with_fit}), which showed a similar reduced $\chi^2$ value to the proper motion maps. This fit again recovered the cluster center and the position angle (compare with \autoref{fig:voronoi_map_gaussian_fit}) albeit with larger statistical errors. Contrary to the velocity dispersion map, a second Gaussian component did not improve the fit (likely due to the larger bin sizes and statistical uncertainties).

\clearpage
\section{Search for metallicity dependent kinematics}
\label{sec:metallicity}
It is well known that \omc{} hosts multiple stellar populations with a wide spread in metallicity (see introduction). Differences in the spatial distribution and the kinematics of the different subpopulations can help to constrain their origin \citep[see][for the case of M54]{2019ApJ...886...57A, 2020ApJ...892...20A,2022ApJ...939..118K}. Various papers have studied the spatial distribution of the different subpopulations in \omc. Early works using ground-based data \citep{2007ApJ...654..915S,2009A&A...507.1393B} found an increase in the concentration of the helium-enhanced blue main sequence within the inner region of the cluster, while at larger radii ($r>10\arcmin$) the ratio of the different helium abundance subpopulations remains approximately constant (however \citealt{2017AJ....153..175C} and \citealt{2020ApJ...891..167C} found evidence for a more extended distribution of the blue main-sequence stars). Most recently, \cite{2024ApJ...970..152N} found no radial variation of the stellar distribution with MUSE-based metallicity measurements within the half-light radius, while \cite{2024A&A...688A.180S} confirmed the trend of a more centrally concentrated blue main sequence from \cite{2009A&A...507.1393B} with precise space-based photometry at larger radii ($\sim1-3 r_{\rm HL}$). 

The kinematics of the different subpopulations have been subject to several investigations: E.g. \cite{2010ApJ...710.1032A} found no kinematic differences between the blue and red-main-sequence populations in the inner region of \omc{}. At larger radii ($\sim3.5\,r_{\rm HL}$), \cite{2018ApJ...853...86B} found significant differences between the anisotropy, the systemic rotation, and the state of energy equipartition for the different populations.

\begin{figure*}[h]
  \centering
    \includegraphics[width=0.45\textwidth]{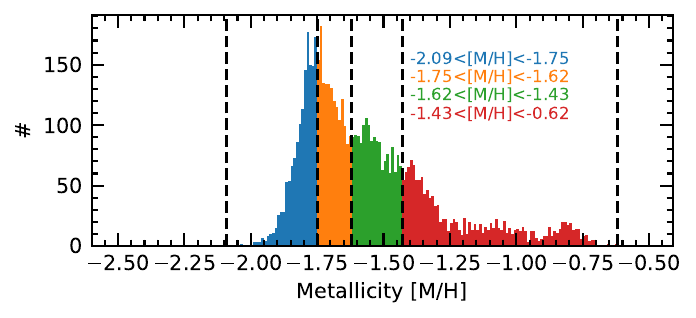}
    \includegraphics[width=0.9\textwidth]{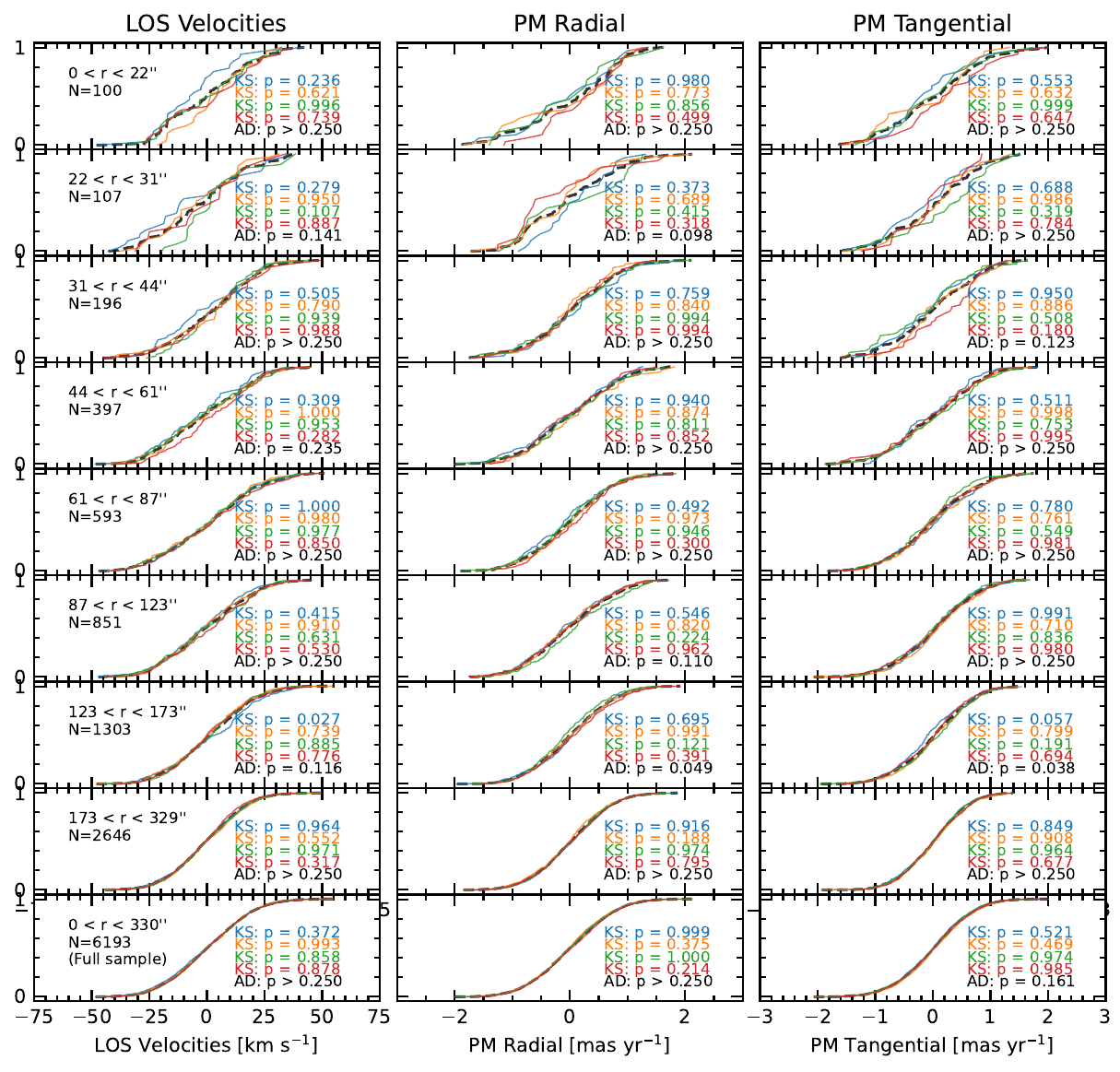}
  \caption{\textit{Top:} Metallicity distribution for bright ($m_{\rm F625W}<17$), well-measured stars in the catalog. We split the sample into metallicity quartiles indicated with different colors. \textit{Bottom:} Normalized cumulative distribution functions for the 3 velocity components and for 8 different radial bins. We compared the velocity distribution for each metallicity quartile with the overall velocity distribution (black dashed line) using a Kolmogorov–Smirnov (KS) and a k-sample Anderson Darling (AD) test but found no significant differences. The $p$ values for the null hypothesis are shown in each panel.}
  \label{fig:metallicity}
\end{figure*}

In this work, we focus on the bright end of the sample, where precise line-of-sight velocities and metallicity measurements are available. To do so, we further restrict the sample of 24,928 stars with a full 3D velocity measurement to $m_{\rm F625W}<17$ (this limit is necessary to obtain reliable and bias-free metallicity, see \citealt{2024ApJ...970..152N}), which leaves us with a subset of 6193 stars. The metallicity distribution for this sample is shown in \autoref{fig:metallicity} (\textit{upper panel}). We then further split the data set into 4 quartiles in metallicity and search for differences in the velocity dispersion in different radial bins (again we used an adaptive logarithmic binning scheme with N$_{min}=100$ and $\Delta$Log$r=0.15$) and for all 3 velocity components (LOS velocity, radial proper motion, tangential proper motion), see \autoref{fig:metallicity} (\textit{lower grid of plots}). In each radial bin and for each velocity component we run a Kolmogorov–Smirnov (KS) and a k-sample Anderson-Darling test to see whether the distribution of an individual metallicity quartile differed significantly from the total distribution. The $p$-values for the null hypothesis (i.e. that the samples are drawn from the same distribution) for both tests are shown in \autoref{fig:metallicity}. We do not find significant ($p < 0.05$) deviations between the distributions in any of the velocity components (radial and tangential proper motion, LOS velocity) and in any of our radial bins.

This is consistent with the picture of well-mixed populations within the half-light radius of \omc{}, see also \cite{2024ApJ...970..152N}. It is also consistent with the recent results of \cite{2025arXiv250217755V}, who found no metallicity-dependent kinematics using both Gaia and the oMEGACat II HST data.
More subtle kinematic differences between the subpopulations may still be discovered when using the full proper motion sample instead of the sample limited to bright stars with a reliable spectroscopic metallicity measurement. This is the subject of a planned future project. 

\clearpage

\section{Energy Equipartition}
\label{sec:ee}

Due to two-body relaxation processes, globular clusters evolve towards a state of energy equipartition, in which lower-mass stars show higher velocity dispersion than stars with higher masses \citep{1969ApJ...158L.139S}.
Traditionally, the state of energy equipartition has been parametrized by fitting the parameter $\eta$, where the velocity dispersion $\sigma$ shows the following dependence on the mass $m$:
\begin{equation}
\label{eq:eeeta}
    \sigma\propto m^{-\eta}.
\end{equation}
A value of $\eta=0.5$ would mean full energy equipartition, however, N-body simulations show that globular clusters only reach partial energy equipartition \citep{2003MNRAS.340..227B,2013MNRAS.435.3272T}.

A different parametrization of the state of energy equipartition using the so-called equipartition mass ($m_{eq}$) was introduced in \cite{2016MNRAS.458.3644B} with the equation:
\begin{equation}
    \sigma (m)=
\begin{cases} \sigma_0\,\text{exp}(-\frac{1}{2}\frac{m}{m_{\rm eq}}) & \text{if}\, m\leq m_{\rm eq}\\ \sigma_{\rm eq}\,(\frac{m}{m_{\rm eq}})^{-1/2} & \text{if}\,m> m_{\rm eq}\end{cases}
\end{equation}
where $\sigma_0$ indicates the velocity dispersion for massless particles and $\sigma_{\rm eq}$ corresponds to the value of velocity dispersion for $m=m_{\rm eq}$. The $m_{\rm eq}$ mass is a proxy for the level of partial energy equipartition reached by the system, such that for $m > m_{\rm eq}$ the system is in full energy equipartition.

The state of energy equipartition is a measure for the overall evolutionary state of the cluster \citep{2016ApJ...827...12B} and other underlying properties such as the presence of black holes,  which can reduce the level of energy equipartition in the luminous stars in the cluster \citep{2018ApJ...864...13W,2023MNRAS.525.3136A,2024MNRAS.529..331D}.
Initial anisotropy in the velocity distribution of a stellar cluster has been shown to influence how fast the system evolved towards energy equipartition \citep{2021MNRAS.504L..12P,2024A&A...689A.313P}, with tangentially anisotropic systems showing a more rapid evolution. \cite{2022MNRAS.509.3815P} further predict differences in the evolution towards energy equipartition for the projected radial and the tangential components of the velocity that can be measured via proper motions.

Observationally, the energy equipartition can be studied by comparing the velocity dispersion measured for stars in different mass bins. However, this is challenging due to the need for velocity measurements with reliable uncertainties over a wide range of stellar masses and magnitudes. While spectroscopic LOS velocity measurements are typically limited to bright evolved stars that have similar masses, HST-based proper motion catalogs have recently enabled the study of energy equipartition in a variety of globular clusters \citep[e.g.][]{2018ApJ...861...99L,2022ApJ...934..150L} including \omc{} \citep{2010ApJ...710.1032A,2018ApJ...853...86B,2022ApJ...936..154W}.
Due to its high precision and depth, the oMEGACat proper motion catalog is perfectly suited to extend these existing studies to lower masses and wider radial coverage.
\subsection{Estimation of Stellar Masses}
\begin{figure*}
  \centering
    \includegraphics[width=0.9\textwidth]{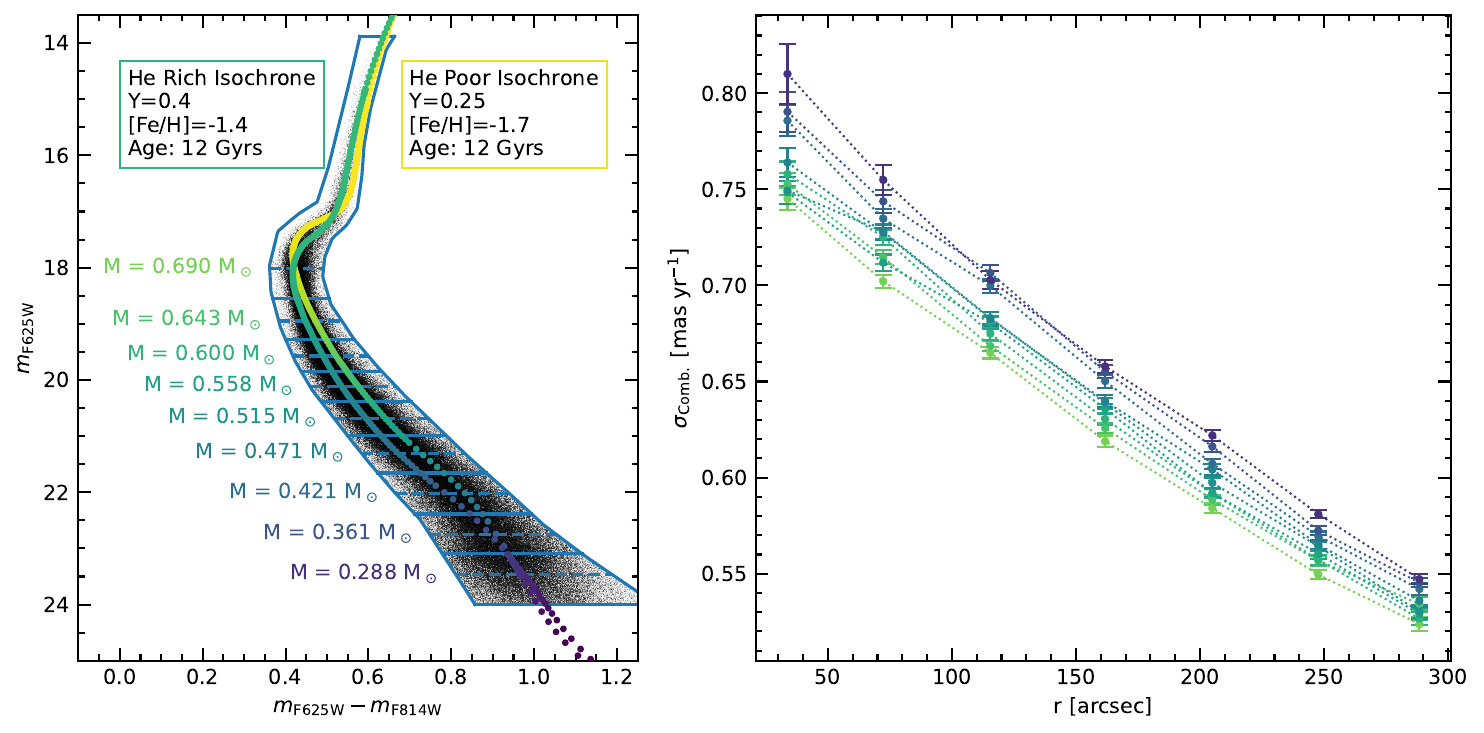}
  \caption{\textit{Left: $m_{\rm F625W}$, $m_{\rm F814W}$} color-magnitude diagram used to split the data set in 10 equally populated $m_{\rm F625W}$ bins. For each bin, a weighted mean mass is estimated using two isochrones to account for both the helium-rich and helium-poor populations in \omc. \textit{Right:} Combined velocity dispersion profile determined for the different mass bins and for 7 different radial bins.}
  \label{fig:eep_mass_estimates}
\end{figure*}

As a first step for our energy equipartition studies, we split our high-quality proper-motion subset into 9 equally populated bins in magnitude (see \autoref{fig:eep_mass_estimates}). Due to the complex stellar populations in \omc{}, it is not straightforward to directly infer stellar masses from their color-magnitude-diagram position. As we focus on the overall kinematics (and do not yet aim to study each subpopulation separately) we use the following approximation (adapted from \citealt{2018ApJ...853...86B}, but with different weights for the different stellar populations) to determine the mean stellar mass in each magnitude bin:
We use two different 12 Gyr isochrones representing the helium-rich and the helium-poor population of the main sequence to infer the mean mass for the different magnitude bins (see \autoref{fig:eep_mass_estimates}, \textit{left}).
For each magnitude bin, we calculated the mean magnitude and interpolated the two isochrones to infer the corresponding stellar mass. We then combined the two different mass estimates, giving them an equal weight (due to their similar fraction within the half-light radius, see \citealt{2009A&A...507.1393B}).
The isochrones were obtained using the Dartmouth Stellar Evolution Database \citep{2007AJ....134..376D,2008ApJS..178...89D}. We used the following parameters: 
\begin{itemize}
    \item Helium Rich Isochrone: Y=0.4,	 [Fe/H] = -1.4 Weight: 50\% 
    \item Helium Poor Isochrone: Y=0.25, [Fe/H] = -1.7 Weight: 50\%
    \item Reddening: E(B-V)=0.16
\end{itemize}
The mean masses per bin in our sample extend from 0.288\,M$_\odot$ to 0.690\,M$_\odot$, a large range that was previously not accessible in the inner regions of \omc{}.

\subsection{Variation of the energy equipartition with radius}
In addition to splitting our data set into different mass bins, we also split it into 7 different annular radial bins (each with a width of 45\arcsec). This allows us to probe the state of energy equipartition at different radii of the cluster. For each magnitude and radial bin we determine the proper-motion dispersion both for the combined proper motion, but also the radial and tangential components individually. A plot showing the different dispersion profiles is shown in \autoref{fig:eep_mass_estimates} \textit{(right)}. Already in this plot, it can be seen, that the high-mass stars show lower velocity dispersion than stars with lower masses, indicative of at least some level of energy equipartition. In the next step, we fit the mass dependency of the velocity dispersion in each radial bin using either the classical $\eta$ or the Bianchini $m_{\rm eq}$ parametrization. The results are shown in \autoref{fig:eep_classic_eta} and \autoref{fig:eep_bianchini}.

We can make the following general observations:
At all radii and for both proper motion directions we can observe at least some degree of energy equipartition. The energy equipartition is highest in the innermost bin ($\eta=0.088\pm0.017$; $m_{\rm eq}$=(2.97$\pm$0.69)\,M$_\odot$) and decreases towards larger radii (at the half-light radius: $\eta=0.049\pm0.009$; $m_{\rm eq}$=(4.51$\pm$0.70)\,M$_\odot$). This trend is consistent with measurements at significantly larger radii ($\eta=0.030\pm0.019$; $r\sim 3.5\,r_{\rm HL}\sim 975\arcsec$), see \cite{2018ApJ...853...86B}.

\subsection{Anisotropy in the energy equipartition}
\begin{figure*}[h]
  \centering
    \includegraphics[width=0.9\textwidth]{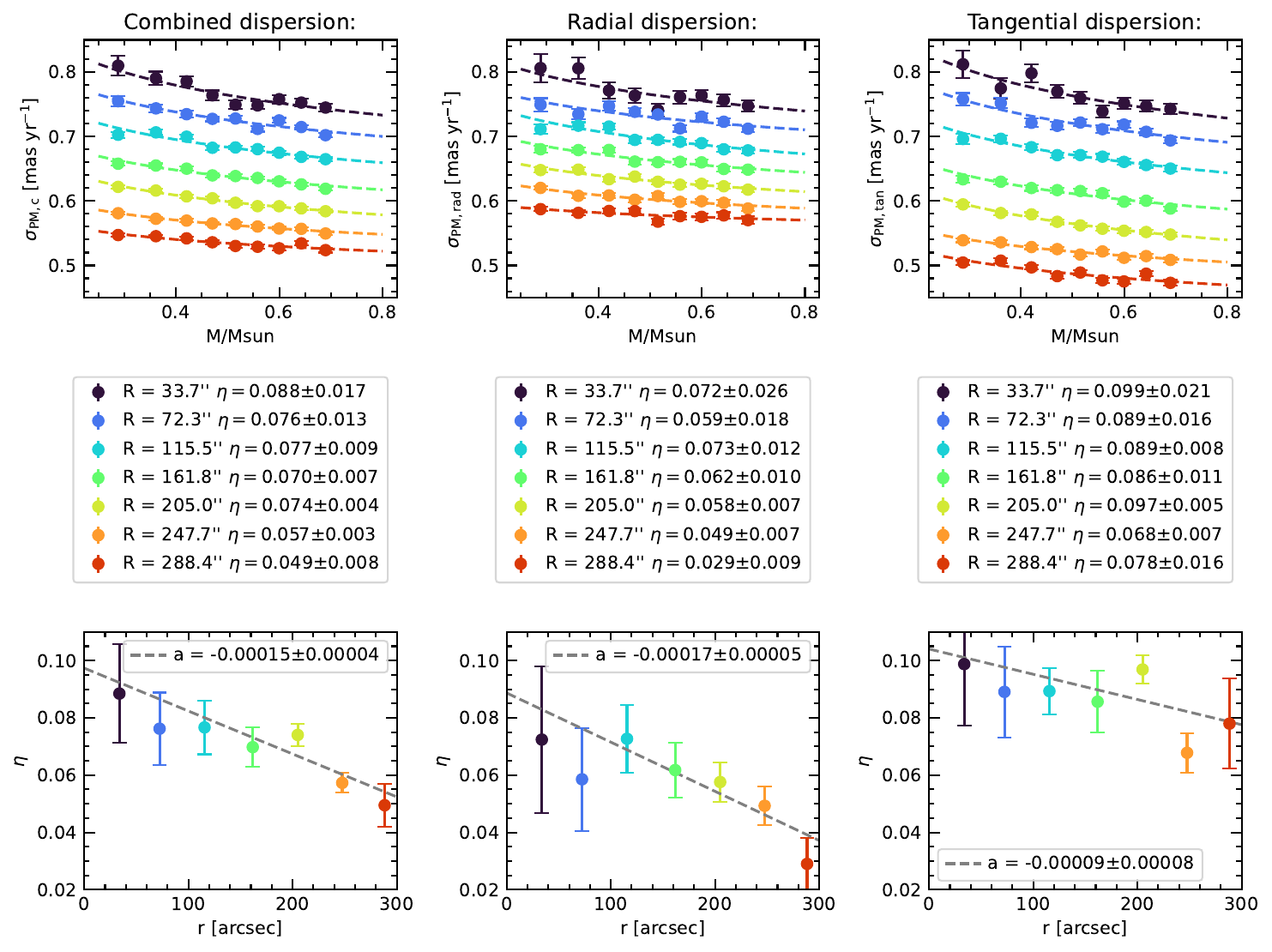}
  \caption{\textit{Top: } Variation of the velocity dispersion with stellar mass. We compare the behavior of the combined velocity dispersion with the individual spatial components (\textit{left:} combined velocity dispersion, \textit{middle: } radial velocity dispersion, \textit{right: } tangential velocity dispersion). The different colors show the measurements in 7 different radial bins. The dashed line shows the best fit of the mass dependence using the classical $\eta$ parametrization. In the middle row, we show the numerical values of the fit results and in the bottom row, we show the radial behavior of the energy equipartition parameter. For both spatial directions, we can observe a decrease in the degree of energy equipartition with radius, indicated by a lower value of the parameter $\eta$. This trend is stronger for the radial component than for the tangential component.}
  \label{fig:eep_classic_eta}
\end{figure*}

\begin{figure*}[h]
  \centering
    \includegraphics[width=0.9\textwidth]{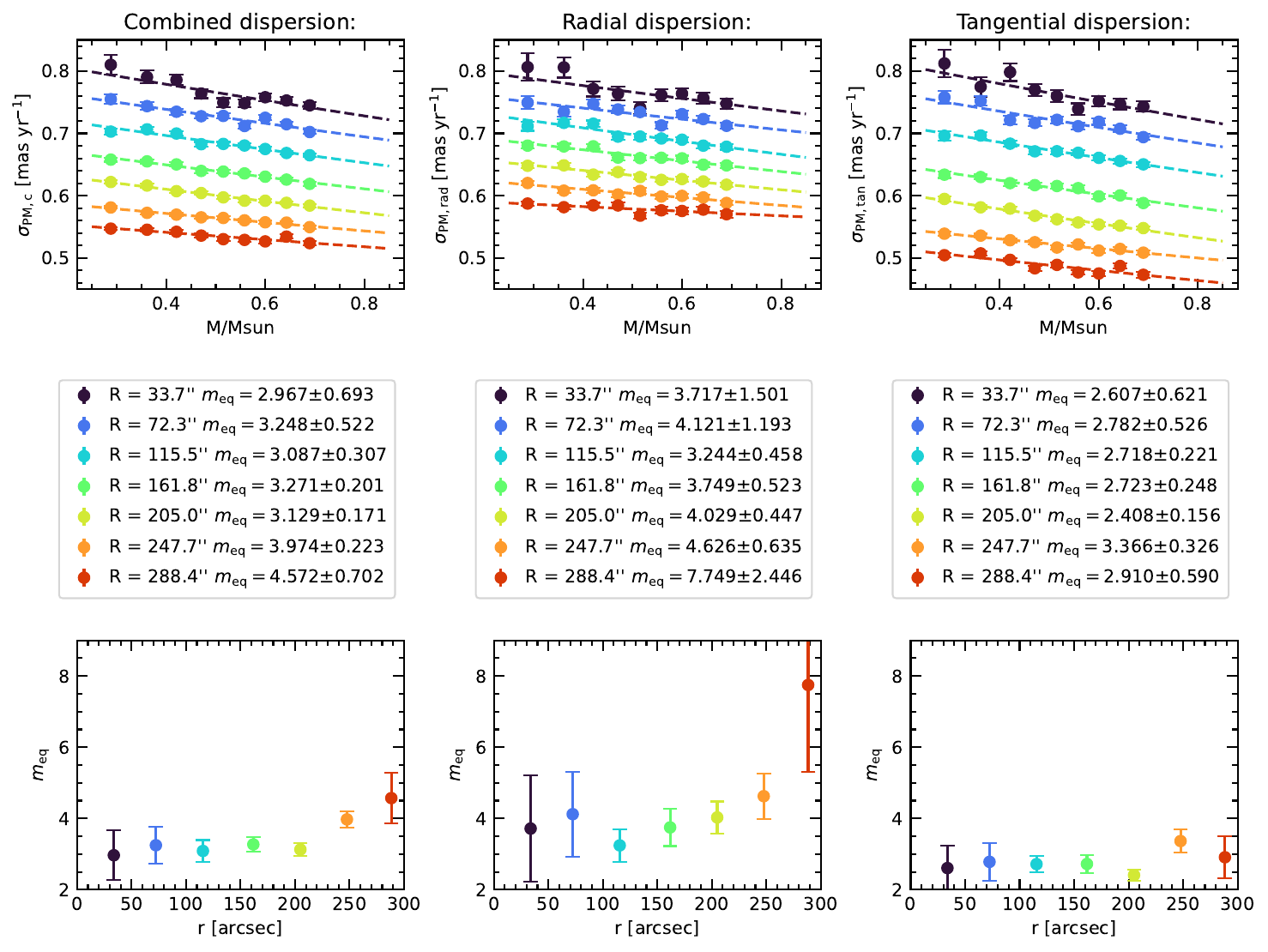}
  \caption{This Figure is equivalent to \autoref{fig:eep_classic_eta} and compares the spatial variability of the degree of energy equipartition for the different components of the proper motion. Instead of the $\eta$ parameter, we use the energy equipartition mass ($m_{\rm eq}$) parametrization \citep{2016MNRAS.458.3644B} to quantify the degree of energy equipartition. A lower value of $m_{\rm eq}$ indicates a higher degree of energy equipartition.}
  \label{fig:eep_bianchini}
\end{figure*}

The overall trends measured for the combined velocity dispersion also hold for the individual radial and tangential directions. However, we can find the following differences between the two directions: While the degree of energy equipartition of the radial component quickly decreases with radius (and the radial velocity dispersion almost does not vary with mass at $r\approx r_{HL}$, see \autoref{fig:eep_classic_eta}), the tangential component shows an overall higher degree of energy equipartition and a weaker dependence with radius. This is tentatively in line with recent simulation results \citep{2021MNRAS.500.2514P,2022MNRAS.509.3815P,2024A&A...689A.313P} that find a faster evolution towards energy equipartition for the tangential component of the velocity.

\clearpage

\section{Summary and Conclusions}
\label{sec:conclusions}
We present a study of several basic kinematic properties of the massive globular cluster \omc{} based on the new spectroscopic and astro-photometric oMEGACat catalogs (\citetalias{2023ApJ...958....8N, 2024ApJ...970..192H}). Due to the enhanced radial coverage and precision of these catalogs and the unique 3-dimensional combination of plane-of-sky proper motions and spectroscopically measured line-of-sight velocities, our analysis significantly improves the kinematic picture of \omc{} and can serve as input for future modeling efforts. We can summarize our analysis as follows:
\begin{itemize}
    \item We determine dispersion profiles with better errors and higher resolution and a range covering 1\arcsec - 300\arcsec, reaching the half-light radius of the cluster. The dispersion profiles show a smooth behavior, with a steady increase from 0.52\,mas\,yr$^{-1}$ (13.6\,km\,s$^{-1}$) at the half-light radius towards $\sim$0.81\,mas\,yr$^{-1}$ (21.1\,km\,s$^{-1}$) within the 10 central arcseconds. We also study the anisotropy of the velocity dispersion field. In the inner region ($r<30\arcsec$), the velocity distribution is isotropic; at larger radii, it starts to become increasingly radially anisotropic, reaching $\sigma_{\rm PM,tan}/\sigma_{\rm PM,rad}=0.0849\pm0.003$ at the half-light radius.
    
    \item Beside the 1-dimensional profiles, we also calculate 2-dimensional velocity dispersion maps. Fitting these maps with a smooth Gaussian model allows us to recover the ellipticity of the velocity field, the position angle, and the kinematic center (which is in agreement with the photometric center derived in \citealt{2010ApJ...710.1032A}).
    
    \item The line-of-sight velocities cover a smaller range in magnitudes but allow us to directly study the rotation curve of the cluster. At small radii ($r\leq$30\arcsec) we can recover the counter-rotating signal described in \citealt{2024MNRAS.528.4941P} (however, with lower significance, as our spectroscopic sample is limited to brighter magnitudes). Outside this region, there is a continuous increase of the rotation curve until it converges to a value of around 7\,km\,s$^{-1}$ at $r\approx$150\arcsec. This is consistent with findings from the plane of sky rotation \citep{2024ApJ...970..192H}. The line-of-sight velocity dispersion profile shows a monotonic increase towards the center. A comparison with the proper motion dispersion profile yields consistent results. By calculating the ratio between the two, we can obtain a kinematic distance estimate of ($5494\pm61$)\,pc, the most precise kinematic distance estimate derived for \omc\ and in good agreement with previous results in \cite{2021MNRAS.505.5957B}.
    \item We split our sample into four metallicity quartiles and searched for variations of the kinematics with metallicity. We do not find metallicity-dependent variations in any of our velocity directions (radial proper motions, tangential proper motions, line-of-sight velocities.).  We note, however, that we had to significantly restrict the data set to a smaller subset of 6193 stars. In principle, the large number of stars with proper motions and precise photometry should enable the search for more subtle variations that were not detectable with the used subset.
    \item The precise proper motion measurements down to faint magnitudes allow us to study the state of energy equipartition of the cluster for stars with a wide range of masses (0.288\,M$_\odot$ to 0.690\,M$_\odot$). We measure a low degree of energy equipartition in the cluster center ($\eta=0.088\pm0.017$) that decreases further towards larger radii. Interestingly, the radial component of the energy equipartition shows a much quicker decrease with radius than the tangential component, whose profile is relatively shallow.
    \item Our kinematic profiles and maps are made public along with the paper in the following Zenodo archive: 
    \href{https://doi.org/10.5281/zenodo.14978551}{DOI:10.5281/zenodo.14978551}\\A description of the data products and tabular versions of the kinematic profiles is given in \autoref{sec:dataproducts}.
\end{itemize}
The next step for a better understanding of the dynamics of \omc{} will be to fit dynamical models to the kinematic data (Pechetti et al. in prep.; Smith et al. in prep.). Our rich data set, which allowed us to accurately measure many peculiar features of \omc{} such as rotation, flattening, anisotropy, partial energy equipartition, and fast-moving central stars, poses both a challenge and an opportunity for all future modeling efforts.
While our data covers the region within the half-light radius, it can be complemented with proper motion measurements obtained with the HST at larger distances from the cluster center \citep{2018ApJ...853...86B,2024A&A...688A.180S} and data from the ESA Gaia mission \citep{2023A&A...680A..35G}, in addition, there are collections of individual radial velocities at large radii obtained with multi-object spectrographs \citep[see][]{2018MNRAS.478.1520B}.

\begin{acknowledgments}
We would like to thank the anonymous referee for the helpful comments, which greatly improved this work. ACS acknowledges support from Hubble Space Telescope grant GO-16777. MH thanks Francisco Aros for useful suggestions that improved the manuscript. AFK acknowledges funding from the Austrian Science Fund (FWF) [grant DOI 10.55776/ESP542]. SKA gratefully acknowledges funding from UKRI through a Future Leaders Fellowship (grant MR/Y034147/1). M.A.C. acknowledges the support from FONDECYT Postdoctorado project No. 3230727.
\end{acknowledgments}

%

\vspace{5mm}


\software{
astropy \citep{2022ApJ...935..167A}, matplotlib \citep{2007CSE.....9...90H}, numpy \citep{2020Natur.585..357H}, scipy \citep{2020NatMe..17..261V}, IPython \citep{2007CSE.....9c..21P}, emcee \citep{2013ascl.soft03002F}, VorBin \citep{2003MNRAS.342..345C}}


\clearpage
\appendix

\section{Data products}
\label{sec:dataproducts}
One of the main purposes of this work is to provide the community with a state-of-the-art kinematic analysis of the inner regions of \omc{} using the novel oMEGACat data, that can be used to dynamically model the cluster. Therefore, we publish the following data-products along with the paper. The data can be accessed using the following Zenodo archive: \href{https://doi.org/10.5281/zenodo.14978551}{DOI:10.5281/zenodo.14978551}\\

All files are provided both as fits tables  (file ending: \texttt{*.fits}) and as machine-readable textfiles (file ending: \texttt{*.dat}). The following files are contained in the archive:
\begin{itemize}
    \item IPython Notebook containing examples on how to read and plot the different data products
    \begin{itemize}
        \item \texttt{data\_products\_demonstration\_and\_test.ipynb}
    \end{itemize}
    \item Proper Motion dispersion profiles (combined, radial, tangential) using different binning schemes
    \begin{itemize}
        \item Adaptive logarithmic bins: \texttt{proper\_motion\_dispersion\_log\_bins.fits} (see also \autoref{tab:dispersionpm})
        \item Linear bins: \texttt{proper\_motion\_dispersion\_lin\_bins.fits}
        \item Equal number bins: \texttt{proper\_motion\_dispersion\_equaln\_bins.fits}
    \end{itemize}
    \item LOS profiles (rotation, dispersion, position angle)
    \begin{itemize}
        \item \texttt{los\_profile.fits} (see also \autoref{tab:dispersionlos})
    \end{itemize}
    \item Kinematic maps
    \begin{itemize}
        \item Proper motion dispersion measurements: \texttt{proper\_motion\_dispersion\_voronoi\_bins.fits}
        \item Line-of-sight mean velocity and disperison measurements:\\\texttt{los\_dispersion\_and\_rotation\_voronoi\_bins.fits}
    \end{itemize}
    \item Data to reproduce the energy equipartition experiments
    \begin{itemize}
        \item Velocity dispersion profiles split into 9 equally populated mass bins:\newline\texttt{energy\_equipartition\_profiles.fits}
        \item A table containing information of the mean mass in each bin:\newline\texttt{energy\_equipartition\_massbins.fits} 
    \end{itemize}
    
    \item A set of selections that can be applied to the catalog to obtain the high-quality subsample that we have used for our analysis:
    \begin{itemize}
        \item \texttt{catalog\_and\_selections.fits}\\
        This file contains the selections and for convenience also various other columns taken from the oMEGACat catalogs. A description of the content is given in \autoref{tab:cat_content}.
    \end{itemize}
\end{itemize}
\nolinenumbers
\begin{table*}[h]
\caption{Content of the catalog and selection file}
\label{tab:cat_content}
\scriptsize
\begin{tabular}{p{3cm}p{8cm}l}
\hline
Column   & Description        & Unit \\ \hline
\texttt{ID}                     & oMEGACat II Identifier; same as in \citetalias{2024ApJ...970..192H} \citep{2024ApJ...970..192H}                                          & -                \\
\texttt{RA}                     & Right Ascension $\alpha$;  \citetalias{2024ApJ...970..192H}                                          & degree                 \\
\texttt{DEC}                    & Declination $\delta$;  \citetalias{2024ApJ...970..192H}                                          & degree                 \\
\texttt{x}                      & x coordinate in pixel based coordinate system;  \citetalias{2024ApJ...970..192H}         & 40~mas ($\sim$ 1 WFC3/UVIS pixel)                 \\
\texttt{y}                      & y coordinate in pixel based coordinate system;  \citetalias{2024ApJ...970..192H}        & 40~mas ($\sim$ 1 WFC3/UVIS pixel)                 \\

\texttt{pmra\_corrected}         & Locally corrected proper motion in R.A. direction;  \citetalias{2024ApJ...970..192H}   & mas~yr$^{-1}$                  \\
\texttt{pmdec\_corrected}        & Locally corrected proper motion in Dec. direction;  \citetalias{2024ApJ...970..192H}   & mas~yr$^{-1}$   \\
\texttt{pmra\_corrected\_err}     & Error on locally corrected proper motion in R.A. direction;  \citetalias{2024ApJ...970..192H}   & mas~yr$^{-1}$                      \\
\texttt{pmdec\_corrected\_err}    & Error on locally corrected proper motion in Dec. direction;  \citetalias{2024ApJ...970..192H}   & mas~yr$^{-1}$                     \\
\texttt{vlos}                                      & Line-of-sight velocity; \citetalias{2023ApJ...958....8N}               & km~s$^{-1}$                     \\
\texttt{vlos\_err}                                  & Error on line-of-sight velocity; \citetalias{2023ApJ...958....8N}               & km~s$^{-1}$                     \\
\texttt{f625w}                                     & Magnitude in the F625W filter (diff. red. corrected);  \citetalias{2024ApJ...970..192H}               & - \\
\texttt{f814w}                                     & Magnitude in the F814W filter (diff. red. corrected);  \citetalias{2024ApJ...970..192H}               & - \\
\texttt{nitschai\_id}            & ID in \textit{oMEGACat I} MUSE spectroscopic catalog \citep{2023ApJ...958....8N}   & -                 \\ 
\texttt{voronoi\_bin\_ids\_pm}                        & Attribution to Voronoi bins in proper motion based kinematic maps               & - \\
\texttt{voronoi\_bin\_ids\_los}                       & Attribution to Voronoi bins in line-of-sight based kinematic maps               & - \\
\texttt{selection\_hq\_f625w}                        & High quality flag for F625W photometry               & - \\
\texttt{selection\_hq\_f814w}                        & High quality flag for F814W photometry               & - \\
\texttt{selection\_hq\_astrometry}                   & High quality flag for astrometry               & - \\
\texttt{selection\_hq\_astrometry \_and\_membership}    & Combined criteria for proper motion quality and cluster membership (CMD and vector-point diagram selections). This is the sample used for the proper motion-based analysis.               & - \\
\texttt{selection\_hq\_los}                          & High quality flag for line-of-sight velocity measurements               & - \\
\texttt{selection\_hq\_pm\_and\_los}                   & Combined proper motion and line-of-sight velocity flag. This is the sample used for the line-of-sight velocity based analysis.               & - \\
\hline

\end{tabular}
\end{table*}

\begin{table*}[]
\caption{Tabular version of proper motion dispersion profile. We also show physical velocity values, converted using our kinematic distance estimate of 5494\,pc.}
\label{tab:dispersionpm}
\footnotesize
\centering
\begin{tabular}{lllllllllll}
\hline
$r_\mathrm{lower}$ & $r_\mathrm{median}$ & $r_\mathrm{upper}$ & $N_\mathrm{Stars}$ & $\sigma_{\rm PM, c}$ & $\sigma_{\rm PM, rad}$ &  $\sigma_{\rm PM, tan}$ & $\sigma_{\rm PM, c}$ & $\sigma_{\rm PM, rad}$ &  $\sigma_{\rm PM, tan}$ & Anisotropy \\\relax
& & & & & & & & & & $\sigma_{\rm rad}/\sigma_{\rm tan}$\\\relax
 [arcsec] & [arcsec] & [arcsec] & - & [mas\,yr$^{-1}$] & [mas\,yr$^{-1}$] & [mas\,yr$^{-1}$] & [km\,s$^{-1}$] & [km\,s$^{-1}$] & [km\,s$^{-1}$] & -\\\hline
  0.00 &   1.82 &   2.68 &    100 & 0.793$^{+0.038}_{-0.037}$ & 0.840$^{+0.060}_{-0.062}$ & 0.755$^{+0.059}_{-0.048}$ & 20.67$^{+1.00}_{-0.96}$ & 21.89$^{+1.56}_{-1.61}$ & 19.68$^{+1.53}_{-1.24}$ & 0.899$\pm$0.091 \\
  2.68 &   3.33 &   3.78 &    100 & 0.812$^{+0.041}_{-0.040}$ & 0.813$^{+0.060}_{-0.060}$ & 0.841$^{+0.066}_{-0.050}$ & 21.15$^{+1.07}_{-1.04}$ & 21.18$^{+1.57}_{-1.55}$ & 21.93$^{+1.71}_{-1.30}$ & 1.035$\pm$0.104 \\
  3.78 &   4.24 &   4.67 &    101 & 0.798$^{+0.042}_{-0.037}$ & 0.818$^{+0.059}_{-0.054}$ & 0.792$^{+0.055}_{-0.050}$ & 20.79$^{+1.09}_{-0.96}$ & 21.30$^{+1.54}_{-1.41}$ & 20.63$^{+1.44}_{-1.31}$ & 0.969$\pm$0.093 \\
  4.67 &   5.00 &   5.34 &    101 & 0.885$^{+0.051}_{-0.039}$ & 0.856$^{+0.066}_{-0.053}$ & 0.925$^{+0.079}_{-0.070}$ & 23.06$^{+1.33}_{-1.02}$ & 22.31$^{+1.73}_{-1.39}$ & 24.09$^{+2.05}_{-1.83}$ & 1.080$\pm$0.115 \\
  5.34 &   5.66 &   5.99 &    109 & 0.851$^{+0.036}_{-0.043}$ & 0.894$^{+0.063}_{-0.060}$ & 0.820$^{+0.058}_{-0.049}$ & 22.18$^{+0.95}_{-1.12}$ & 23.29$^{+1.64}_{-1.56}$ & 21.37$^{+1.51}_{-1.28}$ & 0.917$\pm$0.087 \\
  5.99 &   6.36 &   6.72 &    137 & 0.832$^{+0.040}_{-0.036}$ & 0.821$^{+0.055}_{-0.048}$ & 0.836$^{+0.049}_{-0.044}$ & 21.69$^{+1.03}_{-0.94}$ & 21.40$^{+1.43}_{-1.26}$ & 21.80$^{+1.27}_{-1.15}$ & 1.018$\pm$0.085 \\
  6.72 &   7.13 &   7.54 &    147 & 0.768$^{+0.032}_{-0.026}$ & 0.800$^{+0.046}_{-0.045}$ & 0.734$^{+0.053}_{-0.040}$ & 20.01$^{+0.82}_{-0.69}$ & 20.84$^{+1.21}_{-1.17}$ & 19.14$^{+1.39}_{-1.03}$ & 0.918$\pm$0.078 \\
  7.54 &   8.00 &   8.46 &    204 & 0.792$^{+0.028}_{-0.031}$ & 0.747$^{+0.036}_{-0.037}$ & 0.830$^{+0.044}_{-0.041}$ & 20.63$^{+0.73}_{-0.80}$ & 19.46$^{+0.94}_{-0.97}$ & 21.63$^{+1.15}_{-1.08}$ & 1.112$\pm$0.079 \\
  8.46 &   9.05 &   9.50 &    227 & 0.769$^{+0.024}_{-0.025}$ & 0.775$^{+0.036}_{-0.036}$ & 0.763$^{+0.042}_{-0.032}$ & 20.04$^{+0.63}_{-0.64}$ & 20.20$^{+0.94}_{-0.94}$ & 19.88$^{+1.09}_{-0.83}$ & 0.984$\pm$0.066 \\
  9.50 &  10.04 &  10.65 &    271 & 0.796$^{+0.023}_{-0.021}$ & 0.799$^{+0.043}_{-0.034}$ & 0.796$^{+0.037}_{-0.036}$ & 20.74$^{+0.59}_{-0.56}$ & 20.82$^{+1.13}_{-0.89}$ & 20.75$^{+0.95}_{-0.94}$ & 0.997$\pm$0.066 \\
 10.65 &  11.29 &  11.95 &    363 & 0.763$^{+0.019}_{-0.020}$ & 0.794$^{+0.032}_{-0.029}$ & 0.738$^{+0.030}_{-0.026}$ & 19.90$^{+0.51}_{-0.51}$ & 20.69$^{+0.83}_{-0.76}$ & 19.22$^{+0.78}_{-0.68}$ & 0.929$\pm$0.050 \\
 11.95 &  12.70 &  13.41 &    419 & 0.767$^{+0.020}_{-0.018}$ & 0.757$^{+0.026}_{-0.021}$ & 0.779$^{+0.030}_{-0.026}$ & 20.00$^{+0.53}_{-0.46}$ & 19.72$^{+0.69}_{-0.55}$ & 20.29$^{+0.77}_{-0.68}$ & 1.029$\pm$0.049 \\
 13.41 &  14.26 &  15.05 &    540 & 0.775$^{+0.016}_{-0.016}$ & 0.795$^{+0.024}_{-0.026}$ & 0.768$^{+0.024}_{-0.020}$ & 20.20$^{+0.43}_{-0.43}$ & 20.71$^{+0.63}_{-0.68}$ & 20.01$^{+0.64}_{-0.53}$ & 0.966$\pm$0.041 \\
 15.05 &  16.00 &  16.88 &    720 & 0.775$^{+0.015}_{-0.013}$ & 0.789$^{+0.022}_{-0.020}$ & 0.764$^{+0.017}_{-0.022}$ & 20.20$^{+0.40}_{-0.35}$ & 20.56$^{+0.57}_{-0.52}$ & 19.92$^{+0.44}_{-0.58}$ & 0.969$\pm$0.036 \\
 16.88 &  18.03 &  18.94 &    821 & 0.778$^{+0.015}_{-0.013}$ & 0.789$^{+0.018}_{-0.017}$ & 0.766$^{+0.018}_{-0.017}$ & 20.26$^{+0.38}_{-0.34}$ & 20.57$^{+0.48}_{-0.45}$ & 19.97$^{+0.46}_{-0.45}$ & 0.971$\pm$0.031 \\
 18.94 &  20.16 &  21.25 &   1099 & 0.776$^{+0.012}_{-0.011}$ & 0.783$^{+0.015}_{-0.017}$ & 0.769$^{+0.016}_{-0.014}$ & 20.23$^{+0.30}_{-0.29}$ & 20.41$^{+0.39}_{-0.43}$ & 20.05$^{+0.41}_{-0.36}$ & 0.982$\pm$0.027 \\
 21.25 &  22.62 &  23.85 &   1426 & 0.756$^{+0.010}_{-0.009}$ & 0.763$^{+0.015}_{-0.016}$ & 0.750$^{+0.014}_{-0.015}$ & 19.70$^{+0.25}_{-0.23}$ & 19.87$^{+0.40}_{-0.41}$ & 19.54$^{+0.36}_{-0.39}$ & 0.983$\pm$0.028 \\
 23.85 &  25.28 &  26.76 &   1779 & 0.767$^{+0.010}_{-0.009}$ & 0.771$^{+0.013}_{-0.012}$ & 0.764$^{+0.013}_{-0.012}$ & 19.99$^{+0.25}_{-0.24}$ & 20.10$^{+0.34}_{-0.31}$ & 19.91$^{+0.33}_{-0.33}$ & 0.990$\pm$0.023 \\
 26.76 &  28.43 &  30.02 &   2299 & 0.748$^{+0.008}_{-0.007}$ & 0.738$^{+0.013}_{-0.010}$ & 0.757$^{+0.012}_{-0.011}$ & 19.50$^{+0.20}_{-0.18}$ & 19.22$^{+0.34}_{-0.26}$ & 19.72$^{+0.30}_{-0.28}$ & 1.026$\pm$0.022 \\
 30.02 &  31.95 &  33.68 &   2912 & 0.764$^{+0.008}_{-0.007}$ & 0.761$^{+0.011}_{-0.009}$ & 0.767$^{+0.011}_{-0.009}$ & 19.90$^{+0.21}_{-0.18}$ & 19.82$^{+0.29}_{-0.24}$ & 19.98$^{+0.28}_{-0.24}$ & 1.008$\pm$0.019 \\
 33.68 &  35.81 &  37.79 &   3587 & 0.758$^{+0.006}_{-0.006}$ & 0.754$^{+0.008}_{-0.009}$ & 0.765$^{+0.008}_{-0.010}$ & 19.75$^{+0.17}_{-0.15}$ & 19.65$^{+0.20}_{-0.24}$ & 19.93$^{+0.22}_{-0.26}$ & 1.014$\pm$0.017 \\
 37.79 &  40.17 &  42.40 &   4709 & 0.741$^{+0.005}_{-0.005}$ & 0.745$^{+0.009}_{-0.008}$ & 0.736$^{+0.008}_{-0.007}$ & 19.31$^{+0.13}_{-0.14}$ & 19.40$^{+0.24}_{-0.21}$ & 19.18$^{+0.22}_{-0.17}$ & 0.989$\pm$0.015 \\
 42.40 &  45.04 &  47.58 &   5591 & 0.737$^{+0.005}_{-0.005}$ & 0.744$^{+0.006}_{-0.008}$ & 0.732$^{+0.008}_{-0.007}$ & 19.21$^{+0.14}_{-0.12}$ & 19.40$^{+0.17}_{-0.20}$ & 19.07$^{+0.20}_{-0.17}$ & 0.983$\pm$0.013 \\
 47.58 &  50.58 &  53.38 &   6940 & 0.740$^{+0.005}_{-0.004}$ & 0.745$^{+0.006}_{-0.006}$ & 0.735$^{+0.006}_{-0.006}$ & 19.28$^{+0.12}_{-0.11}$ & 19.42$^{+0.16}_{-0.16}$ & 19.14$^{+0.16}_{-0.16}$ & 0.986$\pm$0.012 \\
 53.38 &  56.70 &  59.89 &   9066 & 0.735$^{+0.004}_{-0.003}$ & 0.740$^{+0.005}_{-0.005}$ & 0.732$^{+0.006}_{-0.006}$ & 19.15$^{+0.11}_{-0.09}$ & 19.29$^{+0.14}_{-0.14}$ & 19.07$^{+0.15}_{-0.15}$ & 0.988$\pm$0.010 \\
 59.89 &  63.58 &  67.20 &  10920 & 0.725$^{+0.003}_{-0.003}$ & 0.731$^{+0.005}_{-0.005}$ & 0.719$^{+0.005}_{-0.005}$ & 18.90$^{+0.09}_{-0.09}$ & 19.04$^{+0.13}_{-0.14}$ & 18.75$^{+0.14}_{-0.12}$ & 0.985$\pm$0.010 \\
 67.20 &  71.42 &  75.40 &  13445 & 0.724$^{+0.003}_{-0.003}$ & 0.728$^{+0.004}_{-0.005}$ & 0.720$^{+0.004}_{-0.004}$ & 18.87$^{+0.07}_{-0.08}$ & 18.96$^{+0.11}_{-0.12}$ & 18.76$^{+0.10}_{-0.11}$ & 0.990$\pm$0.008 \\
 75.40 &  80.09 &  84.60 &  16942 & 0.712$^{+0.003}_{-0.002}$ & 0.721$^{+0.003}_{-0.004}$ & 0.704$^{+0.004}_{-0.004}$ & 18.56$^{+0.07}_{-0.06}$ & 18.80$^{+0.09}_{-0.10}$ & 18.34$^{+0.10}_{-0.10}$ & 0.975$\pm$0.007 \\
 84.60 &  89.93 &  94.93 &  20309 & 0.705$^{+0.002}_{-0.003}$ & 0.711$^{+0.003}_{-0.003}$ & 0.698$^{+0.003}_{-0.003}$ & 18.37$^{+0.06}_{-0.07}$ & 18.54$^{+0.08}_{-0.08}$ & 18.18$^{+0.09}_{-0.08}$ & 0.981$\pm$0.006 \\
 94.93 & 100.73 & 106.51 &  23917 & 0.692$^{+0.002}_{-0.002}$ & 0.702$^{+0.003}_{-0.003}$ & 0.681$^{+0.003}_{-0.003}$ & 18.02$^{+0.05}_{-0.06}$ & 18.29$^{+0.08}_{-0.08}$ & 17.75$^{+0.07}_{-0.08}$ & 0.971$\pm$0.006 \\
106.51 & 112.97 & 119.50 &  27379 & 0.682$^{+0.002}_{-0.002}$ & 0.694$^{+0.003}_{-0.003}$ & 0.671$^{+0.003}_{-0.002}$ & 17.76$^{+0.06}_{-0.04}$ & 18.08$^{+0.08}_{-0.08}$ & 17.48$^{+0.07}_{-0.06}$ & 0.967$\pm$0.006 \\
119.50 & 126.76 & 134.09 &  30196 & 0.669$^{+0.002}_{-0.002}$ & 0.686$^{+0.003}_{-0.003}$ & 0.652$^{+0.003}_{-0.003}$ & 17.44$^{+0.05}_{-0.05}$ & 17.86$^{+0.07}_{-0.07}$ & 17.00$^{+0.07}_{-0.07}$ & 0.952$\pm$0.005 \\
134.09 & 142.43 & 150.45 &  34236 & 0.655$^{+0.002}_{-0.002}$ & 0.678$^{+0.003}_{-0.002}$ & 0.629$^{+0.003}_{-0.002}$ & 17.06$^{+0.04}_{-0.05}$ & 17.68$^{+0.07}_{-0.06}$ & 16.40$^{+0.07}_{-0.06}$ & 0.928$\pm$0.005 \\
150.45 & 160.11 & 168.80 &  43889 & 0.636$^{+0.001}_{-0.001}$ & 0.662$^{+0.002}_{-0.002}$ & 0.610$^{+0.002}_{-0.002}$ & 16.58$^{+0.04}_{-0.04}$ & 17.25$^{+0.06}_{-0.06}$ & 15.89$^{+0.06}_{-0.05}$ & 0.921$\pm$0.004 \\
168.80 & 179.11 & 189.40 &  54408 & 0.621$^{+0.001}_{-0.001}$ & 0.647$^{+0.002}_{-0.002}$ & 0.593$^{+0.002}_{-0.002}$ & 16.18$^{+0.04}_{-0.03}$ & 16.86$^{+0.05}_{-0.06}$ & 15.46$^{+0.05}_{-0.04}$ & 0.917$\pm$0.004 \\
189.40 & 201.15 & 212.51 &  61707 & 0.601$^{+0.001}_{-0.001}$ & 0.634$^{+0.002}_{-0.002}$ & 0.565$^{+0.002}_{-0.002}$ & 15.65$^{+0.03}_{-0.03}$ & 16.51$^{+0.05}_{-0.04}$ & 14.73$^{+0.05}_{-0.04}$ & 0.892$\pm$0.004 \\
212.51 & 225.54 & 238.44 &  71611 & 0.584$^{+0.001}_{-0.001}$ & 0.619$^{+0.002}_{-0.002}$ & 0.546$^{+0.001}_{-0.002}$ & 15.21$^{+0.03}_{-0.03}$ & 16.13$^{+0.04}_{-0.04}$ & 14.23$^{+0.04}_{-0.04}$ & 0.882$\pm$0.003 \\
238.44 & 252.36 & 267.54 &  78997 & 0.562$^{+0.001}_{-0.001}$ & 0.602$^{+0.002}_{-0.002}$ & 0.519$^{+0.001}_{-0.001}$ & 14.63$^{+0.02}_{-0.03}$ & 15.68$^{+0.04}_{-0.04}$ & 13.52$^{+0.04}_{-0.03}$ & 0.862$\pm$0.003 \\
267.54 & 280.94 & 300.18 &  60637 & 0.544$^{+0.001}_{-0.001}$ & 0.586$^{+0.002}_{-0.001}$ & 0.497$^{+0.001}_{-0.001}$ & 14.17$^{+0.03}_{-0.03}$ & 15.26$^{+0.05}_{-0.04}$ & 12.96$^{+0.03}_{-0.04}$ & 0.849$\pm$0.003 \\
300.18 & 311.12 & 346.07 &  18485 & 0.520$^{+0.002}_{-0.002}$ & 0.562$^{+0.003}_{-0.003}$ & 0.474$^{+0.003}_{-0.002}$ & 13.54$^{+0.05}_{-0.05}$ & 14.66$^{+0.07}_{-0.08}$ & 12.36$^{+0.07}_{-0.06}$ & 0.843$\pm$0.006 \\

\hline
 \end{tabular}
\end{table*}

\begin{table*}[]
\caption{Tabular version of line-of-sight (LOS) dispersion profile.}
\label{tab:dispersionlos}
\footnotesize
\centering
\begin{tabular}{lllllll}
\hline
$r_\mathrm{lower}$ & $r_\mathrm{median}$ & $r_\mathrm{upper}$ & $N_\mathrm{Stars}$ & $v_{\rm LOS}$ & $\sigma_{\rm LOS}$ &  $\theta_0$ \\\relax
 [arcsec] & [arcsec] & [arcsec] & - &  [km\,s$^{-1}$] & [km\,s$^{-1}$] & [degree]\\\hline
 0.00 &   5.66 &   7.90 &     50 & 10.3$^{+5.1}_{-4.8}$ & 20.9$^{+2.5}_{-1.9}$ & -135.8$^{+21.2}_{-20.9}$ \\
  7.90 &   9.41 &  11.35 &     50 & 4.7$^{+3.3}_{-3.9}$ & 19.0$^{+2.1}_{-2.0}$ & -169.7$^{+50.5}_{-46.4}$ \\
 11.35 &  13.41 &  14.51 &     50 & 8.3$^{+4.0}_{-4.0}$ & 21.0$^{+2.3}_{-1.9}$ & 175.6$^{+26.9}_{-32.4}$ \\
 14.51 &  16.09 &  17.08 &     51 & 10.0$^{+4.4}_{-4.7}$ & 20.4$^{+2.6}_{-1.7}$ & 41.2$^{+21.7}_{-23.9}$ \\
 17.08 &  18.68 &  19.68 &     51 & 8.2$^{+3.9}_{-3.7}$ & 17.9$^{+1.8}_{-1.7}$ & 116.9$^{+24.4}_{-29.1}$ \\
 19.68 &  20.91 &  22.09 &     64 & 5.9$^{+2.9}_{-3.3}$ & 18.0$^{+1.7}_{-1.6}$ & 74.6$^{+35.2}_{-32.6}$ \\
 22.09 &  23.55 &  24.78 &     92 & 0.8$^{+2.6}_{-3.1}$ & 20.9$^{+2.1}_{-1.7}$ & -12.1$^{+99.1}_{-113.5}$ \\
 24.78 &  26.31 &  27.79 &    137 & 3.1$^{+2.9}_{-2.5}$ & 21.4$^{+1.5}_{-1.3}$ & 36.9$^{+45.3}_{-50.7}$ \\
 27.79 &  29.52 &  31.17 &    119 & 3.2$^{+2.4}_{-2.5}$ & 18.4$^{+1.4}_{-1.2}$ & 79.5$^{+42.9}_{-37.9}$ \\
 31.17 &  33.24 &  34.98 &    179 & 1.6$^{+2.0}_{-1.9}$ & 18.4$^{+0.9}_{-0.8}$ & -128.4$^{+59.4}_{-72.9}$ \\
 34.98 &  37.15 &  39.26 &    224 & 4.6$^{+1.9}_{-1.4}$ & 18.3$^{+0.9}_{-0.9}$ & 116.4$^{+20.7}_{-21.8}$ \\
 39.26 &  41.55 &  44.05 &    282 & 2.5$^{+1.5}_{-1.4}$ & 19.1$^{+0.9}_{-0.8}$ & 73.8$^{+37.1}_{-38.4}$ \\
 44.05 &  46.97 &  49.43 &    359 & 3.8$^{+1.2}_{-1.4}$ & 18.6$^{+0.6}_{-0.6}$ & 101.7$^{+25.0}_{-21.7}$ \\
 49.43 &  52.60 &  55.47 &    457 & 3.9$^{+1.2}_{-1.1}$ & 17.6$^{+0.5}_{-0.6}$ & 122.4$^{+16.1}_{-16.5}$ \\
 55.47 &  58.79 &  62.26 &    590 & 2.6$^{+1.0}_{-1.2}$ & 19.6$^{+0.6}_{-0.6}$ & 71.0$^{+26.2}_{-27.7}$ \\
 62.26 &  66.08 &  69.86 &    591 & 4.4$^{+1.0}_{-1.0}$ & 18.4$^{+0.6}_{-0.6}$ & 115.8$^{+12.6}_{-16.5}$ \\
 69.86 &  74.29 &  78.39 &    766 & 4.3$^{+0.9}_{-0.8}$ & 18.4$^{+0.5}_{-0.5}$ & 86.1$^{+13.2}_{-12.5}$ \\
 78.39 &  83.43 &  87.95 &    903 & 4.4$^{+0.8}_{-0.8}$ & 18.8$^{+0.5}_{-0.4}$ & 92.1$^{+11.3}_{-14.7}$ \\
 87.95 &  93.34 &  98.68 &   1107 & 4.3$^{+0.7}_{-0.7}$ & 17.2$^{+0.4}_{-0.3}$ & 99.5$^{+9.4}_{-11.4}$ \\
 98.68 & 104.57 & 110.72 &   1291 & 5.0$^{+0.7}_{-0.8}$ & 17.8$^{+0.3}_{-0.4}$ & 107.6$^{+8.1}_{-8.8}$ \\
110.72 & 117.26 & 124.24 &   1239 & 4.8$^{+0.8}_{-0.7}$ & 17.1$^{+0.3}_{-0.3}$ & 92.1$^{+8.5}_{-9.5}$ \\
124.24 & 131.59 & 139.40 &   1274 & 5.4$^{+0.6}_{-0.7}$ & 17.3$^{+0.4}_{-0.4}$ & 94.8$^{+7.4}_{-7.3}$ \\
139.40 & 148.36 & 156.42 &   1755 & 6.6$^{+0.5}_{-0.4}$ & 16.4$^{+0.3}_{-0.3}$ & 98.3$^{+5.4}_{-5.1}$ \\
156.42 & 166.17 & 175.50 &   2098 & 6.8$^{+0.5}_{-0.5}$ & 15.9$^{+0.2}_{-0.2}$ & 110.0$^{+4.3}_{-4.6}$ \\
175.50 & 186.00 & 196.92 &   2358 & 7.0$^{+0.5}_{-0.5}$ & 15.5$^{+0.2}_{-0.3}$ & 109.4$^{+3.4}_{-3.5}$ \\
196.92 & 208.94 & 220.95 &   2453 & 6.4$^{+0.4}_{-0.4}$ & 14.6$^{+0.2}_{-0.2}$ & 100.9$^{+4.0}_{-3.6}$ \\
220.95 & 234.05 & 247.90 &   2662 & 6.1$^{+0.4}_{-0.4}$ & 14.2$^{+0.2}_{-0.2}$ & 103.4$^{+3.9}_{-4.0}$ \\
247.90 & 261.11 & 278.14 &   2401 & 6.6$^{+0.3}_{-0.3}$ & 13.7$^{+0.2}_{-0.2}$ & 105.9$^{+3.8}_{-3.7}$ \\
278.14 & 290.28 & 332.28 &   1275 & 6.6$^{+0.5}_{-0.4}$ & 12.5$^{+0.2}_{-0.3}$ & 106.8$^{+5.4}_{-4.1}$ \\
\hline
\end{tabular}

\end{table*}
\nolinenumbers

\section{Plots describing the selections}
\begin{figure*}[h]
  \centering
    \includegraphics[width=0.9\textwidth]{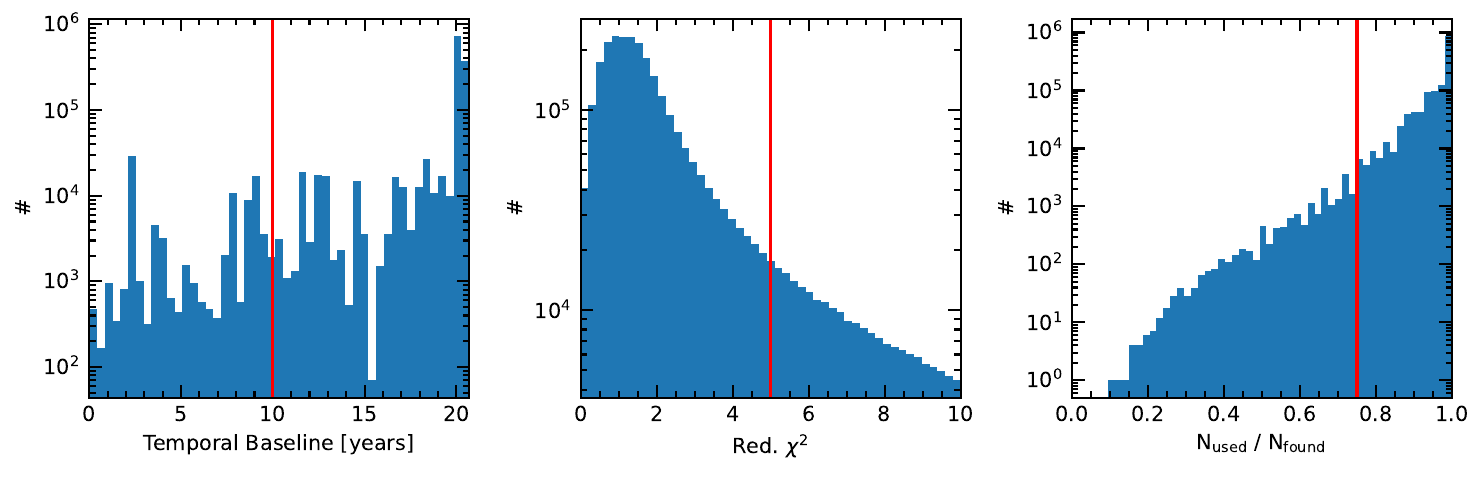}
  \caption{Histograms and thresholds of the astrometric quality parameters used to define the high-quality subsample of reliable proper motion measurements. \textit{Left: } Temporal baseline used for proper motion determination, \textit{Center:} Reduced $\chi^2$ value of the linear fit to the astrometric data used to determine the proper motions. \textit{Right: } Fraction of used measurements for the proper motion fit. A low value indicates unreliable astrometry.}
  \label{fig:selectioncs_global}
\end{figure*}

\begin{figure*}[h]
  \centering
    \includegraphics[width=0.9\textwidth]{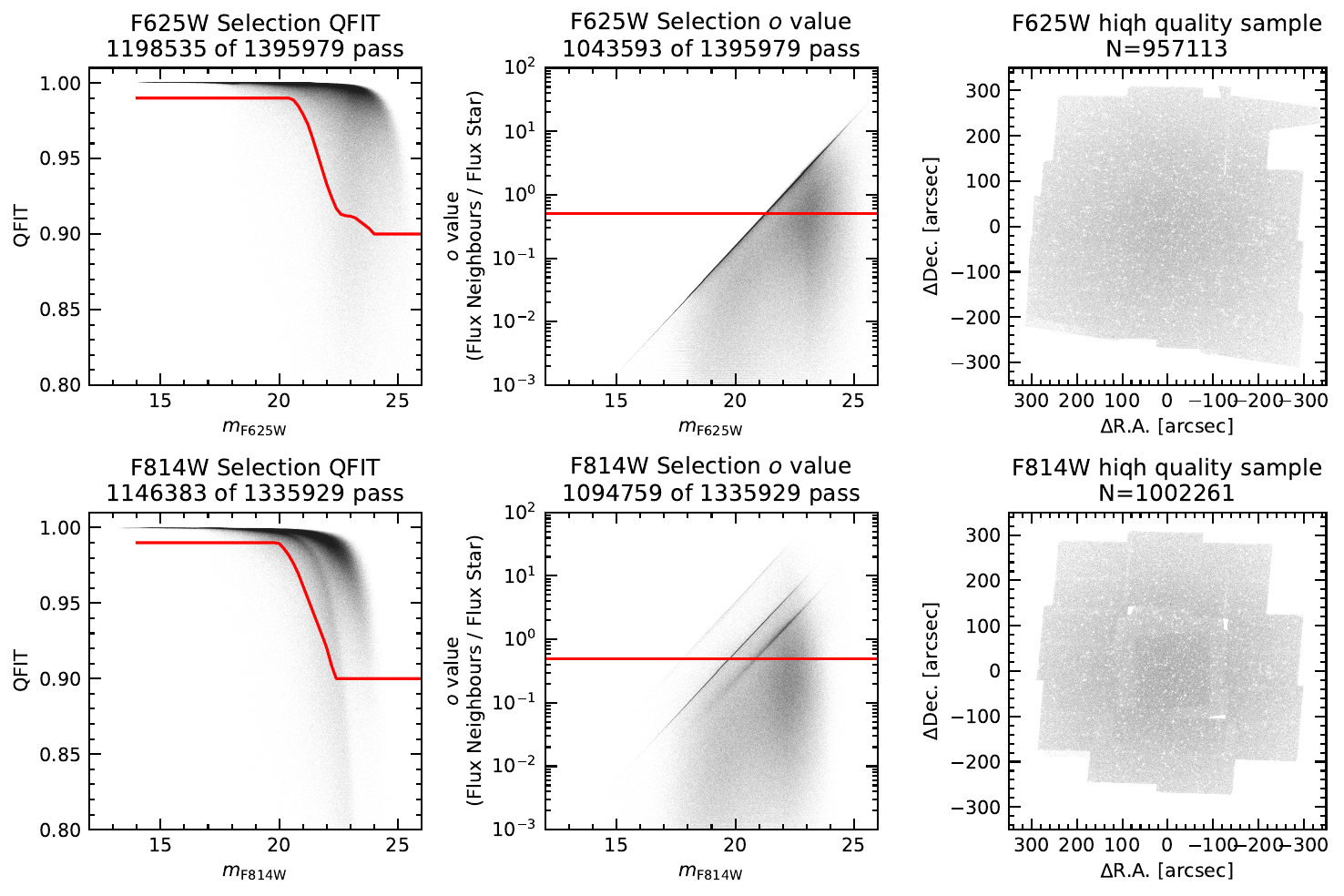}
  \caption{Photometric quality selections used to determine a high-quality subset of the data for both the ACS WFC F625W filter (\textit{top}) and the WFC3/UVIS F814W filter (\textit{bottom}). \textit{Left:} Magnitude dependent threshold on the QFIT value that characterizes how well the point-spread-function could be fit to the data. \textit{Center: }$o$ value that characterized the fraction of flux from neighboring sources for each photometric measurement. \textit{Right: } Spatial distribution of well-measured stars. The measurements in the F625W are quite uniformly distributed. The F814W measurements show some spatial dependences and minor gaps, due to the distribution of pointings.}
  \label{fig:selectioncs_phot}
\end{figure*}

Several cuts in astrometric and photometric quality parameters were used to restrict the full oMEGACat II catalog to a subsample of reliable proper motion measurements (see also \autoref{sec:selections}). \autoref{fig:selectioncs_global} shows the overall distributions and thresholds for the parameters used for the astrometric selection. \autoref{fig:selectioncs_phot} shows magnitude-dependent photometric quantities (quality-of-fit parameter QFIT,  relative flux value of neighboring sources) and the spatial distribution of photometrically well-measured sources.

\section{Comparison of kinematic distance estimates}
Our new kinematic distance estimate for \omc{} is $d=(5494\pm61)$\,pc. In \autoref{tab:distances} we compare it with several other recent distance estimates, for which there is overall good agreement. We refer to \cite{2021MNRAS.505.5957B} for a more extensive discussion of previous distance measurements to \omc{}.
\begin{table*}[]
\centering
\caption{Comparison of our new kinematic distance estimate with various recent literature estimates.}
\label{tab:distances}
\footnotesize
\begin{tabular}{lll}
\hline
Distance & Data and Method & Reference   \\ \hline
\textbf{($5494\pm61$})\,pc& \textbf{oMEGACat, kin. dist.} &\textbf{ This work} \\
(5240$\pm$110)\,pc& Gaia EDR3, parallax & \cite{2021ApJ...908L...5S} \\
(5485$^{+0.302}_{-0.272}$)\,pc& Gaia EDR3, parallax & \cite{2021MNRAS.505.5957B} \\
(5359$\pm$141)\,pc& Gaia EDR3, kin. dist. & \cite{2021MNRAS.505.5957B} \\
(5264$\pm$121)\,pc& HST, kin. dist. & \cite{2021MNRAS.505.5957B} \\

 \hline
\end{tabular}
\end{table*}

\section{Variation of binning schemes}
\label{app:binning}
\subsection{Radii of circular bins}
\begin{figure*}[h]
  \centering
    \includegraphics[width=0.9\textwidth]{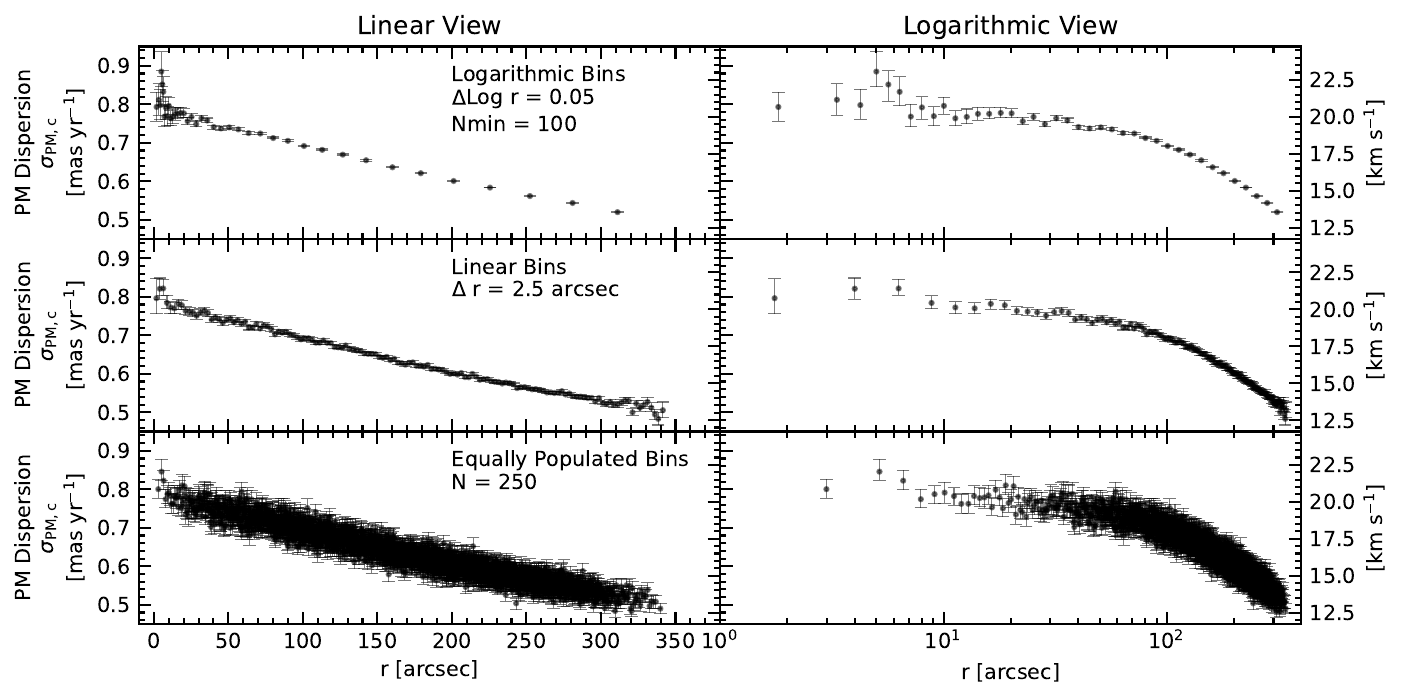}
  \caption{Comparison of the total proper motion dispersion profile determined with various binning schemes (\textit{top:} adaptive logarithmic, \textit{middle:} linear/equal radius, \textit{bottom:} equi-populated). The profiles are shown in both linear (\textit{left}) and logarithmic (\textit{right}) scale. The different profiles show overall agreement but differ in resolution and scatter.}
  \label{fig:binning_comparison}
\end{figure*}

Traditionally, the velocity dispersion has been measured in circular radial bins. The choice of the bin radii is ultimately arbitrary, and with more than 600,000 individual stellar measurements and a large radial range of around 350\arcsec\ several binning schemes are feasible. In \autoref{fig:binning_comparison} we explore three different binning options: our first binning choice is an adaptive, logarithmic binning scheme. The stepsize is $\Delta\text{log}\,r=0.05$, but we require a minimum number of at least 100 stars per bin. This binning choice has a high resolution in the centermost region while being more coarse in the outer regions. The second explored binning scheme is a simple linear scheme with a bin size of $\Delta r=2.5\arcsec$, naturally maintaining a uniform resolution in both the inner and outer regions. Finally, we explored equally populated bins with a number of $N=250$ stars per bin. The advantage of this scheme is the uniform uncertainties in all bins. However, the resolution in the center is comparatively low, while the bin density is very large at larger radii. The three binning schemes agree with each other within their uncertainties and we make all profiles publicly available so that the user can choose the scheme most appropriate to their science case. For our further discussions, we use the first adaptive logarithmic binning scheme.
\subsection{Testing elliptical instead of circular bins}
\label{app:ellipcitalbins}
\begin{figure*}[h]
  \centering
    \includegraphics[width=0.9\textwidth]{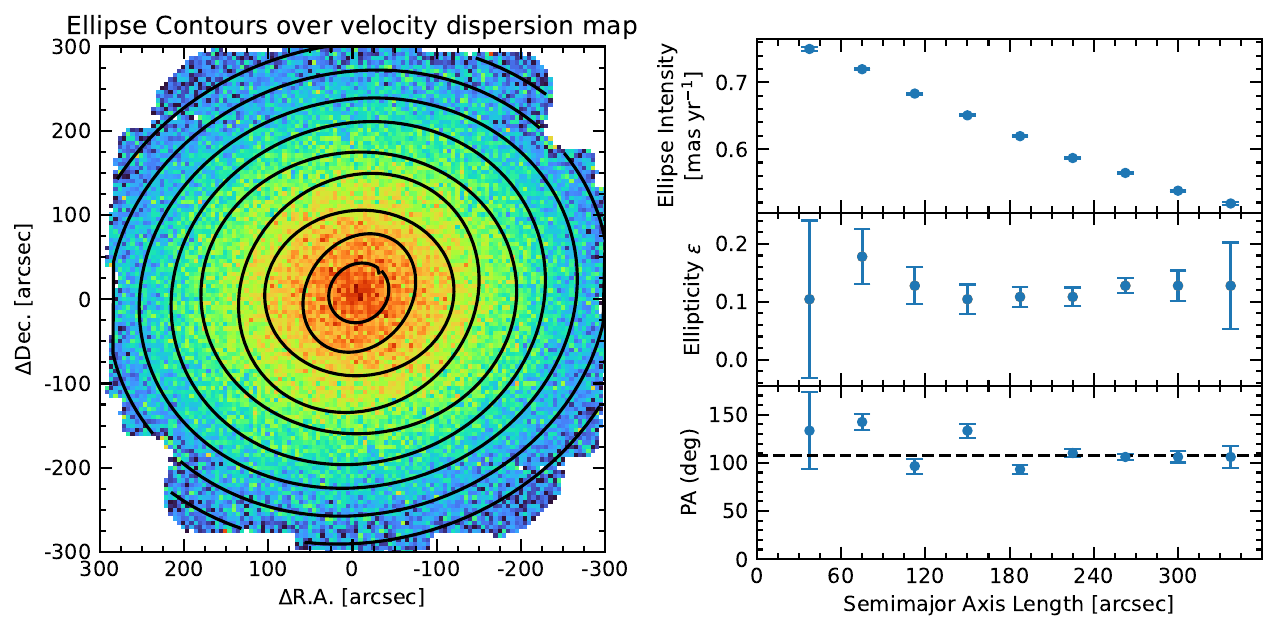}
  \caption{Determination of the ellipticity of the velocity field. \textit{Left: } A symmetrized map of the proper motion dispersion determined in $5\arcsec\times5\arcsec$ grid cells. The black ellipses show the isodispersion contours fit with \texttt{photutils.isophote}. \textit{Right: } Profiles of the determined ellipse parameters of the isodisperion contours (\textit{top:} dispersion, \textit{middle:} ellipticty, \textit{bottom:} position angle). The median ellipticity is $\epsilon_{\mathrm{disp.}}=0.12$, the position angle $PA=108^\circ$.}
  \label{fig:ellipticity}
\end{figure*}

The stellar density and the surface-brightness of \omc{} show significant flattening with variable ellipticity that reaches a maximum of $\epsilon=1-\frac{b}{a}=0.16$ at $r=8\arcmin$ and a mean value of $\epsilon=0.10$ \citep{1983A&A...125..359G,2003MNRAS.345..683P,2020ApJ...891..167C}. Therefore, the choice of circular bins might not fully capture the nature of the dispersion profile of the cluster. To determine the ellipticity of the 2D velocity dispersion field we first calculated a dispersion map on a regular grid with a bin size of $5\arcsec\times5\arcsec$. We then symmetrized the map using the photometric \cite{2010ApJ...710.1032A} center as a pivot point to fill in the gaps in the data set (see \autoref{fig:ellipticity}, \textit{left}). Then we used the \texttt{photutils.isophot} function to fit elliptical ``isophots'' (or isodispersion contours) to the map. At smaller radii ($r\leq2\arcmin$), the ellipticity and position angle are poorly constrained and show large scatter. At larger radii, the ellipticity converges to a median value of $\epsilon_{\mathrm{disp.}}=0.12$ with a median position angle of $PA=108^\circ$ in good agreement with the light distribution and the results of the Gaussian fits (see \autoref{fig:voronoi_map_gaussian_fit}). Using these values we calculated the dispersion profile using elliptical bins instead of circular bins but did not find significant differences when comparing bins with the same mean radii.


\subsection{Variation of 2D binning schemes}
\subsubsection{Variation of the target number of stars in Voronoi bins}
\begin{figure*}[h]
  \centering
    \includegraphics[width=0.9\textwidth]{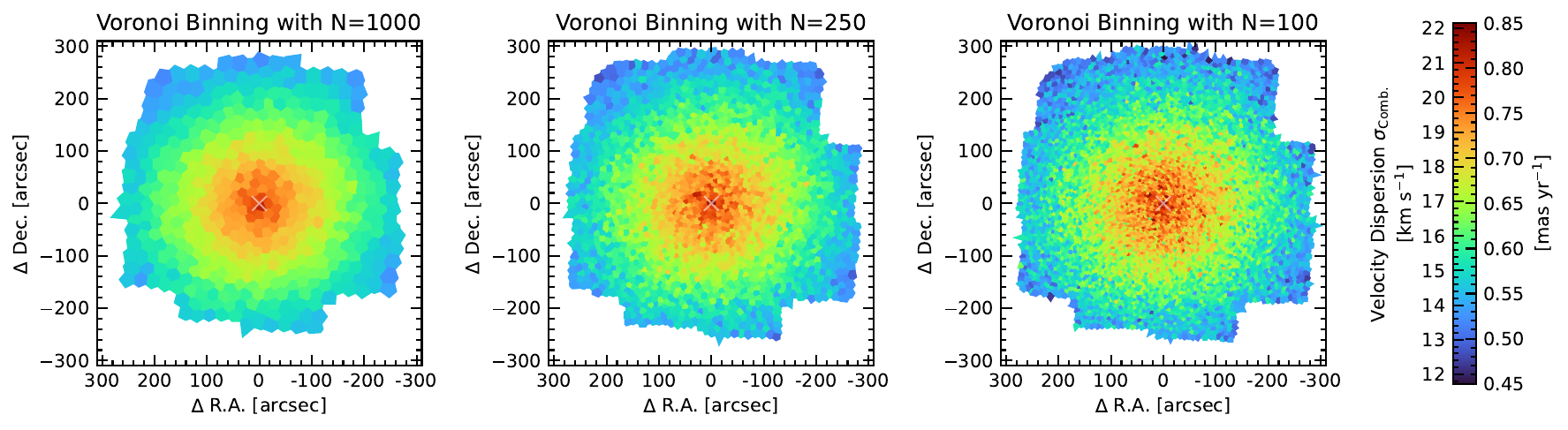}
  \caption{Comparison of proper motion dispersion maps using three different Voronoi binning schemes with a different number of measurements per bin. \textit{Left: } $N=1000$, \textit{Center: } $N=250$, \textit{Right:} $N=100$}
  \label{fig:voronoi_maps_binning_variation}
\end{figure*}

Our main kinematic maps (see \autoref{fig:voronoi_map_proper_motion}) were determined using a Voronoi binning scheme with a target number of $N=250$ stars per bin. In \autoref{fig:voronoi_maps_binning_variation} we also show kinematic maps with $N=1000$ and $N=100$. This comparison demonstrates that $N=250$ is a good compromise between spatial resolution and noise in the individual bins. With $N=1000$ the variation due to stochastic noise is lower than the spatial variation, meaning that information is lost due to the large bin size. With $N=100$, the stochastic noise starts to dominate, meaning that no further information is retained by a finer bin size.
\subsubsection{Comparison between Voronoi binning and nearest neighbor schemes}
\begin{figure*}[h]
  \centering
    \includegraphics[width=0.9\textwidth]{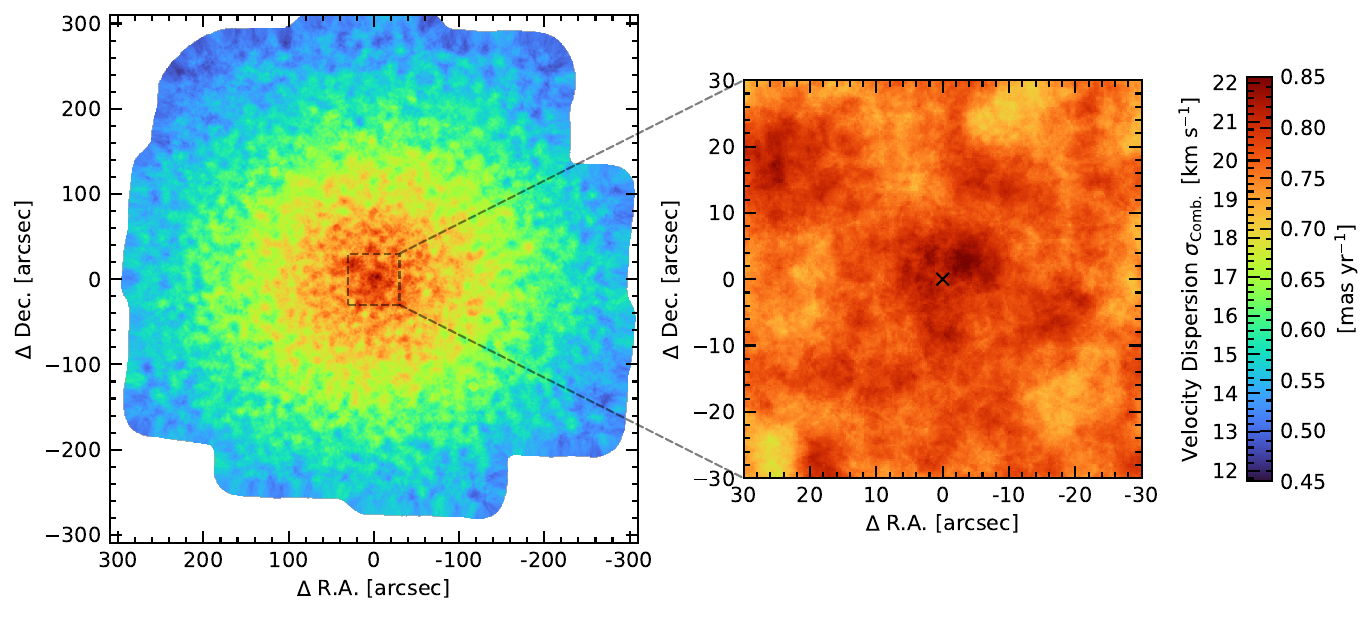}
  \caption{A proper motion dispersion map based on a nearest neighbor binning scheme with $N=250$. This Figure allows to compare the KNN scheme with the Vornoi binning scheme used in \autoref{fig:voronoi_map_proper_motion}.}
  \label{fig:knn_250_map}
\end{figure*}

The employed Voronoi binning schemes offer the advantage of splitting the data into fully independent bins that each contain a similar number of stars and, therefore, yield a similar statistical noise level. Another commonly used method (see, e.g., \citealt{2024MNRAS.528.4941P,2024ApJ...970..152N}) to derive 2D binned maps in stellar fields is to use a nearest-neighbor scheme to group the stars. In \autoref{fig:knn_250_map} we used this scheme to create a kinematic map with the same properties as in \autoref{fig:voronoi_map_proper_motion}. Overall there is good agreement between the two spatial binning methods. The KNN map shows granularity with a feature size comparable to the size of the Voronoi bins in \autoref{fig:voronoi_map_proper_motion}, which is set by the search radius necessary to find the required number of neighbors. At small scales, the KNN map has a smoother appearance (as neighboring points share a large part of their star sample); however, this should not be mistaken as better precision. It just means that the uncertainties are correlated on scales smaller than the neighbor search radius.


\bibliography{kinematics_bib,additional_bib}{}
\bibliographystyle{aasjournal}


\end{CJK*}
\end{document}